\documentclass[12pt]{article}
 \usepackage{amssymb}
 \makeatletter
\renewcommand\thesection{\@Roman\c@section}
\renewcommand\thesubsection{\thesection.\@arabic\c@subsection}
\makeatother

\newcommand{\sect}[1]{\setcounter{equation}{0}\section{#1}}

\begin{document}

\newfont{\elevenmib}{cmmib10 scaled\magstep1}
\newcommand{\preprint}{
   %\begin{flushleft}
    % \elevenmib Yukawa\, Institute\, Kyoto\\
   %\end{flushleft}\vspace{-1.3cm}
   \begin{flushright}\normalsize
     {\tt hep-th/0503003} \\ March 2005
   \end{flushright}}

\begin{center}
{\Large \bf Drinfeld Twists and Symmetric Bethe Vectors of
Supersymmetric Fermion Models }

 \vskip.2in {\large Shao-You
Zhao$^{1,2}$, Wen-Li Yang$^{1,3}$ and Yao-Zhong Zhang$^{1}$
 } \vskip.2in
 {\em $^{1}$ Department of Mathematics, University of
Queensland, Brisbane 4072, Australia }\\

 {\em $^{2}$  Department of Physics, Beijing Institute of
Technology, Beijing 100081, China }\\

 {\em $^{3}$ Institute of Modern Physics, Northwest University,
Xi'an 710069, China}\\

\vskip.1in {\em E-mail:} syz@maths.uq.edu.au,
wenli@maths.uq.edu.au, yzz@maths.uq.edu.au

\end{center}
\begin{abstract}
We construct the Drinfeld twists (factorizing $F$-matrices) of the
$gl(m|n)$-invariant  fermion model. Completely symmetric
representation of the pseudo-particle creation operators of the
model are obtained in the basis provided by the $F$-matrix (the
$F$-basis). We resolve the hierarchy of the nested Bethe vectors
in the $F$-basis for the $gl(m|n)$ supersymmetric model.
\end{abstract}
%\newpage
\sect{Introduction}

In \cite{Maillet96}, it was realized that the $R$-matrices for the
one-dimensional integrable XXX  and XXZ spin chain systems are
factorized in terms of certain non-degenerate lower-triangular
$F$-matrices (Drinfeld twists \cite{Drinfeld83})
\begin{equation}
R_{12}(u_1,u_2)=F^{-1}_{21}(u_2,u_1)F_{12}(u_1,u_2).
\label{de:R-FF}
\end{equation}
This leads to the natural $F$-basis for the analysis of these
models. Working in the $F$-basis, the pseudo-particle creation
operators of the systems take the completely symmetric form.
Compared to the original Bethe vectors of the models, Bethe
vectors in the $F$-basis are dramatically simplified and can be
written down explicitly. These results allow form factors,
correlation functions and spontaneous magnetizations of the
systems to be represented in exact and compact form
\cite{Kitanine98,Izergin98}.

The results of \cite{Maillet96} were generalized to other models
including the models associated with any finite-dimensional
irreducible representations of the Yangian $Y[gl(2)]$
\cite{Terras99}, the $gl(n)$ rational Heisenberg model
\cite{Albert00}, the elliptic XYZ and Belavin models
\cite{Albert0005,Albert0007}.

There are integrable models which do not come from a bosonic
algebra. Examples include the Perk-Schultz model \cite{Perk81},
whose exact solution and the Bethe ansatz equations were obtained
in \cite{Schultz83,Vega91,Lopes92}. The actual algebra underlying
the Perk-Schultz model is the Lie superalgebra $gl(m|n)$: using
the transformation introduced in \cite{Kulish80}, the $R$-matrix
of the Perk-Schultz model is related to the $gl(m|n)$-invariant
$R$-matrix. Integrable models with $gl(m|n)$ supersymmetry are
physically important because they give strongly correlated fermion
models of superconductivity. Interesting examples include the
$gl(2|1)$-invariant supersymmetric $t$-$J$ model, the
$gl(2|2)$-invariant electronic model \cite{Ess9211} and the
supersymmetric $U$ model \cite{Zhang95}, proposed in an attempt to
understanding high-$T_c$ superconductivities. The application of
the hierarchy of the algebraic Bethe ansatz to spin systems
related to Lie superalgebras was given in \cite{Kulish85}.

Recently \cite{zsy04,zsy05}, we have successfully constructed the
Drinfeld twists for the $t$-$J$ models with $gl(2|1)$ and
$U_q(gl(2|1))$ supersymmetries, and resolved the hierarchies of
the nested Bethe vectors of the two models. In this paper, we
construct the factorizing $F$-matrices of the $gl(m|n)$-invariant
fermion model. Working in the $F$-basis, we obtain the symmetric
representations of the monodromy matrix and the creation
operators. Moreover we resolve the hierarchy of the nested Bethe
vectors in the $F$-basis for the $gl(m|n)$ model.

The present paper is organized as follows. In section II, we
introduce some basic notation on the $gl(m|n)$-invariant fermion
model. In section III, we construct the $F$-matrix and its
inverse. In section IV, we give the representation of the
monodromy matrix and creation operators in the $F$-basis. In
section V, specializing to the $gl(2|2)$ invariant superconductive
electronic model, we resolve its nested Bethe vectors in the
$F$-basis. In section VI, we resolve the hierarchy of the Bethe
vectors of the $gl(m|n)$ model in the $F$-basis. We conclude the
paper by offering some discussions in section VII.

\sect{Basic definitions and notation}
 ~~~
Let $V$ be a $(m+n)$-dimensional $gl(m|n)$-module and $R\in
End(V\otimes V)$ be the $R$-matrix associated with this module.
$V$ is $Z_2$-graded, and in the following we choose the following
grading for $V$:  $[1]=\ldots=[m]=1,[m+1]=\ldots=[m+n]=0$. The
graded permutation operator ${\cal P}$ is defined by
\begin{eqnarray}
 {\cal P}(a\otimes b)=(-1)^{[a][b]}(b\otimes a)
\end{eqnarray} or in matrix form
$$({\cal P})^{cd}_{ab}=(-1)^{[c][d]}\delta_{ad}\delta_{bc}.$$
The $R$-matrix depends on the difference of two spectral
parameters $u_1$ and $u_2$ associated with the two copies of $V$,
and is given by
\begin{eqnarray}
 R_{12}(u_1,u_2)=R_{12}(u_1-u_2)%\nonumber\\
% &=& \nonumber\\
          &=&a_{12}I+b_{12}{\cal P}, \label{de:R} \nonumber\\
\end{eqnarray}
where $I$ is the identity operator, and
\begin{eqnarray}
&& a_{12}=a(u_1,u_2)\equiv {u_1-u_2\over u_1-u_2+\eta},\quad \quad
b_{12}=b(u_1,u_2)\equiv{\eta\over u_1-u_2+\eta}
\end{eqnarray}
with $\eta\in C$ being the crossing parameter. One can easily
check that the $R$-matrix satisfies the unitary relation
\begin{equation}
R_{21}R_{12}=1.
\end{equation}
Here and throughout $R_{12}\equiv R_{12}(u_1,u_2)$ and
$R_{21}\equiv R_{21}(u_2,u_1)$. The $R$-matrix satisfies the
graded Yang-Baxter equation (GYBE) \cite{Kulish80}
\begin{equation}
R_{12}R_{13}R_{23}=R_{23}R_{13}R_{12}.
\end{equation}
In terms of the matrix elements defined by
\begin{equation}
R(u)(v^{i'}\otimes v^{j'})=\sum_{i,j}R(u)^{i'j'}_{ij}(v^{i}\otimes
v^{j}),
\end{equation}
the GYBE reads \cite{Kulish80}
\begin{eqnarray}
&&
\sum_{i',j',k'}R(u_1-u_2)^{i'j'}_{ij}R(u_1-u_3)^{i''k'}_{i'k}R(u_2-u_3)^{j''k''}_{j'k'}
    (-1)^{[j']([i']+[i''])}\nonumber\\
&=&\sum_{i',j',k'}R(u_2-u_3)^{j'k'}_{jk}R(u_1-u_3)^{i'k''}_{ik'}R(u_1-u_2)^{i''j''}_{i'j'}
    (-1)^{[j']([i]+[i'])}.
\end{eqnarray}

The quantum monodromy matrix $T(u)$ on a lattice of length $N$ is
defined as
\begin{eqnarray}
T(u)=R_{0N}(u,z_N)R_{0N-1}(u,z_{N-1})_{\ldots} R_{01}(u,z_1),
\label{de:T}
\end{eqnarray}
where the index 0 refers to the auxiliary space  and $\{z_i\}$ are
arbitrary inhomogeneous parameters depending on site $i$. $T(u)$
can be represented in the auxiliary space as a $(m+n)\times (m+n)$
matrix whose elements are operators acting on the quantum space
$V^{\otimes N}$:
\begin{eqnarray}
 {T}(u)=\left(\begin{array}{cccc}
{ A}_{11}(u)& \cdots& {A}_{1\ m+n-1}(u)&{ B}_1(u)\\
 \vdots & \vdots & \vdots&\vdots\\
{ A}_{m+n-1\ 1}(u)& \cdots& { A}_{m+n-1\ m+n-1}(u)&{ B}_{m+n-1}(u)\\
{ C}_{1}(u)& \cdots&{ C}_{m+n-1}(u)&{ D}(u)
 \end{array}\right). \label{de:T-marix}
 \end{eqnarray}

 By using the GYBE, one may prove that the monodromy matrix
satisfies the GYBE
\begin{eqnarray}
R_{12}(u-v)T_1(u)T_2(v)=T_2(v)T_1(u)R_{12}(u-v).\label{eq:GYBE}
\end{eqnarray}
%or in matrix form,
%\begin{eqnarray}
%&&\sum_{i',j'}R(u-v)^{i'j'}_{ij}T(u)^{i''}_{i'}T(v)^{j''}_{j'}(-1)^{[i'']([j']+[j''])}
%    \nonumber\\ && \mbox{} \quad\quad
%=\sum_{i',j'}T(v)^{j'}_{j}T(u)^{i'}_{i}R(u-v)^{i''j''}_{i'j'}(-1)^{[i]([j]+[j'])}.
%\end{eqnarray}

Define the transfer matrix $t(u)$
\begin{eqnarray}
t(u)=str_0T(u),\label{de:t}
\end{eqnarray}
where $str_0$ denotes the supertrace over the auxiliary space.
Then the Hamiltonian of the $gl(m|n)$ supersymmetric model is
given by
\begin{equation}
H={d\ln t(u)\over du}|_{u=0}. \label{de:H}
\end{equation}
This model is integrable thanks to the commutativity of the
transfer matrix for different parameters,
\begin{equation}
[t(u),t(v)]=0,
\end{equation}
which can be verified by using the GYBE.

Following \cite{Maillet96}, we now introduce the notation
$R^\sigma_{1\ldots N}$, where $\sigma$ is any element of the
permutation group ${\cal S}_N$. We note that we may rewrite the
GYBE as
\begin{eqnarray}
R_{23}^{\sigma_{23}}T_{0,23}=T_{0,32}R_{23}^{\sigma_{23}},
    \label{eq:RT-sigma23}
\end{eqnarray}
where $\sigma_{23}$ is the transposition of space labels (2,3),
$R_{23}^{\sigma_{23}}\equiv R_{23}$, and $T_{0,23}\equiv
R_{03}R_{02}$. It follows that $R_{1\ldots N}^{\sigma}$ is a
product of elementary $R$-matrices, corresponding to a
decomposition of $\sigma$ into elementary transpositions. With the
help of the GYBE, one may generalize (\ref{eq:RT-sigma23}) to a
$N$-fold tensor product of spaces
\begin{eqnarray}
 R_{1\ldots N}^{\sigma}T_{0,1\ldots N}
  =T_{0,\sigma(1\ldots N)}R_{1\ldots N}^{\sigma},
     \label{eq:RT-sigma}
\end{eqnarray}
where $T_{0,1\ldots N}\equiv R_{0N}\ldots R_{01}.$ This implies
the ``decomposition" law
\begin{eqnarray}
 R^{\sigma'\sigma}_{1\ldots N}
 =R^{\sigma}_{\sigma'(1\ldots N)}
  R^{\sigma'}_{1\ldots N},\label{eq:R-RR}
\end{eqnarray}
for a product of two elements in ${\cal S}_N$. We remark here that
for the elementary transposition $\sigma_{i,i+1}$, $R_{1\ldots
N}^{\sigma_{i,i+1}}=R_{i,i+1}$, and for any $\sigma\in {\cal
S}_N$, $R_{1\ldots N}^{\sigma}$ can be obtained with the help of
(\ref{eq:R-RR}).

Note that $R^{\sigma}_{\sigma'(1\ldots N)}$ satisfies the relation
\begin{eqnarray}
 R^{\sigma}_{\sigma'(1\ldots N)}
 T_{0,\sigma'(1\ldots N)}
 =T_{0,\sigma'\sigma(1\ldots N)}
  R^{\sigma}_{\sigma'(1\ldots N)}. \label{eq:RT-sigma'}
\end{eqnarray}
As in \cite{Albert00}, we write the elements of $R_{1\ldots
N}^{\sigma}$ as
\begin{eqnarray}
 \left(R_{1\ldots N}^{\sigma}\right)
  ^{\alpha_{\sigma(N)}\ldots \alpha_{\sigma(1)}}
  _{\beta_N\ldots \beta_1},
\end{eqnarray}
where the labels in the upper indices are permuted relative to the
lower indices according to $\sigma$.

\sect{$F$-matrices for the $gl(m|n)$ supersymmetric model}

In this section, we construct the factoring $F$-matrix and its
inverse for the supersymmetric $gl(m|n)$ model.

\subsection{The $F$-matrix}
 ~~~~
 For the $R$-matrix (\ref{de:R}), we define the
$F$-matrix
 %%%%%%%%%%%%%%%%%%%%%%%%%%%%%%%%%%%%%%%%%%%%%%%%%%%%%%%%%%%%%%%
%\begin{eqnarray}
% &&F_{12}%\nonumber\\ &=&
%         =\left(\begin{array}{ccccccccc}
% 1&0&0 & 0&0&0 & 0&0&0\\
% 0&1&0 & 0&0&0 & 0&0&0 \\
% 0&0&1 & 0&0&0  & 0&0&0\\
% 0&b_{12}&0 & a_{12}&0&0  & 0&0&0\\
% 0&0&0      & 0&1&0  & 0&0&0\\
% 0&0&0      & 0&0&1  & 0&0&0\\
% 0&0&b_{12} & 0&0&0       & a_{12}&0&0\\
% 0&0&0      & 0&0&b_{12}  & 0&a_{12}&0\\
% 0&0&0      & 0&0&0       & 0&0&1+c_{12}
% \end{array} \right).% \nonumber\\
%% \end{eqnarray}
%It is convenient to write the $F$-matrix as the form
\begin{eqnarray}
F_{12}=\sum_{m+n\geq\alpha_2\geq\alpha_1}
      P_1^{\alpha_1}P_2^{\alpha_2}
      +c_{12}\sum_{\gamma=1}^m P_1^{\gamma}P_2^{\gamma}
      +\sum_{m+n\geq\alpha_1>\alpha_2}
      P_1^{\alpha_1}P_2^{\alpha_2}R_{12},\label{de:F12}
\end{eqnarray}
where $(P_i^\alpha)_{k}^{l}=\delta_{k,\alpha}\delta_{l,\alpha}$ is
the projector acting on $i$th space, and $c_{12}\equiv
a_{12}-b_{12}$. Then by the $R$-matrix (\ref{de:R}) and $F$-matrix
(\ref{de:F12}), we have
\begin{eqnarray}
 F_{21}R_{12}%=\nonumber\\
&=&\left(\sum_{m+n\geq\alpha_1\geq\alpha_2}
      P_2^{\alpha_2}P_1^{\alpha_1}
      +c_{21}\sum_{\gamma=1}^m P_2^{\gamma}P_1^{\gamma}
      +\sum_{m+n\geq\alpha_2>\alpha_1}
      P_2^{\alpha_2}P_1^{\alpha_1}R_{21}\right)R_{12}
      \nonumber\\
&=&\left(\sum_{m+n\geq\alpha_1>\alpha_2}
      P_2^{\alpha_2}P_1^{\alpha_1}
      +(1+c_{21})\sum_{\gamma=1}^m P_2^{\gamma}P_1^{\gamma}
      +\sum_{\gamma=m+1}^{m+n} P_2^{\gamma}P_1^{\gamma}
        \right. \nonumber\\&& \mbox{} \left. ~~~~~~~~~~~~
      +\sum_{m+n\geq\alpha_2>\alpha_1}
      P_2^{\alpha_2}P_1^{\alpha_1}R_{21}\right)R_{12}
      \nonumber\\
&=&\sum_{m+n\geq\alpha_1>\alpha_2}
      P_2^{\alpha_2}P_1^{\alpha_1}R_{12}
      +(1+c_{12})\sum_{\gamma=1}^m P_2^{\gamma}P_1^{\gamma}
      +\sum_{\gamma=m+1}^{m+n} P_2^{\gamma}P_1^{\gamma}
      \nonumber\\&& \mbox{} ~~~~~~~~~~~~
      +\sum_{m+n\geq\alpha_2>\alpha_1}
      P_2^{\alpha_2}P_1^{\alpha_1}
      \nonumber\\
&=&\sum_{m+n\geq\alpha_1>\alpha_2}
      P_2^{\alpha_2}P_1^{\alpha_1}R_{12}
      +c_{12}\sum_{\gamma=1}^m P_2^{\gamma}P_1^{\gamma}
      +\sum_{m+n\geq\alpha_2\geq\alpha_1}
      P_2^{\alpha_2}P_1^{\alpha_1}
      \nonumber\\
&=&F_{12}.
\end{eqnarray}
Here we have used $R_{12}R_{21}=1$ and $c_{12}c_{21}=1$. Some
remarks are in order. The solutions to (\ref{de:R-FF}), i.e. the
$F$-matrices satisfying (\ref{de:R-FF}), are not unique
\cite{Maillet96,Albert00}. In this paper, we only consider
particular solution of the form (\ref{de:F12}), which is
lower-triangle.

We now generalize the $F$-matrix to the $N$-site problem. As is
pointed out in \cite{Albert00}, the generalized $F$-matrix should
satisfy three properties: i) lower-triangularity; ii)
non-degeneracy and
\begin{eqnarray}
\mbox{iii)}\ \ F_{\sigma(1\ldots
N)}(z_{\sigma(1)},\ldots,z_{\sigma(N)})
  R_{1\ldots N}^\sigma(z_1,\ldots,z_N)
 =F_{1\ldots N}(z_1,\ldots,z_N), \label{eq:R-F-N}
\end{eqnarray}
where $\sigma\in {\cal S}_N$ and $z_i$, $i=1,\ldots,N$, are
generic inhomogeneous parameters .

Define the $N$-site $F$-matrix:
\begin{eqnarray}
F_{1,\ldots N}=\sum_{\sigma\in {\cal S}_N}
   \sum_{\alpha_{\sigma(1)}\ldots\alpha_{\sigma(N)}}^{\quad\quad *}
   \prod_{j=1}^N P_{\sigma(j)}^{\alpha_{\sigma(j)}}
   S(c,\sigma,\alpha_\sigma)R_{1\ldots N}^\sigma, \label{de:F}
\end{eqnarray}
where the sum $\sum^*$  is over all non-decreasing sequences of
the labels $\alpha_{\sigma(i)}$:
\begin{eqnarray}
&& \alpha_{\sigma(i+1)}\geq \alpha_{\sigma(i)}\quad \mbox{if}\quad
              \sigma(i+1)>\sigma(i) \nonumber\\
&& \alpha_{\sigma(i+1)}> \alpha_{\sigma(i)}\quad \mbox{if}\quad
              \sigma(i+1)<\sigma(i) \label{cond:F}
\end{eqnarray}
and the c-number function $S(c,\sigma,\alpha_\sigma)$ is given by
\begin{eqnarray}
S(c,\sigma,\alpha_\sigma)\equiv
\exp\{\sum_{l>k=1}^N\delta^{[\gamma]}_{\alpha_{\sigma(k)},\alpha_{\sigma(l)}}
    \ln(1+c_{\sigma(k)\sigma(l)})\}
\end{eqnarray}
with $\gamma=1,\ldots,m$,
$\delta^{[\gamma]}_{\alpha_{\sigma(k)},\alpha_{\sigma(l)}}=1$ for
$\alpha_{\sigma(k)}=\alpha_{\sigma(l)}=\gamma$, and
$\delta^{[\gamma]}_{\alpha_{\sigma(k)},\alpha_{\sigma(l)}}=0$
otherwise.

 %Note that $R^{\sigma}$ is factorized according to $R$'s of
 % the elementary transpositions.
The definition of $F_{1\ldots N}$, (\ref{de:F}), and the summation
condition (\ref{cond:F}) imply that $F_{1\ldots N}$ is a
lower-triangular matrix. Moreover, one can easily check that the
$F$-matrix is non-degenerate because all diagonal elements are
non-zero.

We now prove that the $F$-matrix (\ref{de:F}) satisfies the
property iii). Any given permutation $\sigma\in {\cal S}_N$ can be
decomposed into elementary transpositions of the group ${\cal
S}_N$ as $\sigma=\sigma_1\ldots \sigma_k$ with $\sigma_i$ denoting
the elementary permutation $(i,i+1)$. By (\ref{eq:R-RR}), we have
if the property iii) holds for elementary transposition
$\sigma_i$,
\begin{eqnarray}
&&F_{\sigma(1\ldots N)}R^{\sigma}_{1\ldots N}= \nonumber\\
 &=& F_{\sigma_1\ldots\sigma_k(1\ldots N)}
     R^{\sigma_k}_{\sigma_1\ldots\sigma_{k-1}(1\ldots N)}
     R^{\sigma_{k-1}}_{\sigma_1\ldots\sigma_{k-2}(1\ldots N)}
     \ldots
     R^{\sigma_1}_{1\ldots N}\nonumber\\
 &=&F_{\sigma_1\ldots\sigma_{k-1}(1\ldots N)}
     R^{\sigma_{k-1}}_{\sigma_1\ldots\sigma_{k-2}(1\ldots N)}
     \ldots
     R^{\sigma_1}_{1\ldots N}\nonumber\\
  &=&\ldots
     =F_{\sigma_1(1\ldots N)}R^{\sigma_1}_{1\ldots N}=F_{1\ldots N}.
\end{eqnarray}

For the elementary transposition $\sigma_i$, we have
\begin{eqnarray}
 &&F_{\sigma_i(1\ldots N)}R^{\sigma_i}_{1\ldots N}= \nonumber\\
 &=&\sum_{\sigma\in {\cal S}_N}
   \sum_{\alpha_{\sigma_i\sigma(1)}\ldots\alpha_{\sigma_i\sigma(N)}}^{\quad\quad *}
   \prod_{j=1}^N P_{\sigma_i\sigma(j)}^{\alpha_{\sigma_i\sigma(j)}}
 %  \nonumber\\ &&\times
  S(c,\sigma_i\sigma,\alpha_{\sigma_i\sigma})R_{\sigma_i(1\ldots N)}^\sigma
   R_{1\ldots N}^{\sigma_i} \nonumber\\
 &=&\sum_{\sigma\in {\cal S}_N}
   \sum_{\alpha_{\sigma_i\sigma(1)}\ldots\alpha_{\sigma_i\sigma(N)}}^{\quad\quad *}
   \prod_{j=1}^N P_{\sigma_i\sigma(j)}^{\alpha_{\sigma_i\sigma(j)}}
%   \nonumber\\ &&\times
   S(c,\sigma_i\sigma,\alpha_{\sigma_i\sigma})
   R_{1\ldots N}^{\sigma_i\sigma} \nonumber\\
 &=&\sum_{\tilde\sigma\in {\cal S}_N}
   \sum_{\alpha_{\tilde\sigma(1)}\ldots\alpha_{\tilde\sigma(N)}}^{\quad\quad *(i)}
   \prod_{j=1}^N P_{\tilde\sigma(j)}^{\alpha_{\tilde\sigma(j)}}
   S(c,\tilde\sigma,\alpha_{\tilde\sigma})R_{1\ldots N}^{\tilde\sigma}, \label{eq:FR-F}\nonumber\\
\end{eqnarray}
where $\tilde\sigma=\sigma_i\sigma$, and the summation sequences
of $\alpha_{\tilde\sigma}$ in ${\sum^*}^{(i)}$ now has the form
\begin{eqnarray}
&& \alpha_{\tilde\sigma(j+1)}\geq \alpha_{\tilde\sigma(j)}\quad
\mbox{if}\quad
              \sigma_i\tilde\sigma(j+1)>\sigma_i\tilde\sigma(j), \nonumber\\
&& \alpha_{\tilde\sigma(j+1)}> \alpha_{\tilde\sigma(j)}\quad
\mbox{if}\quad
              \sigma_i\tilde\sigma(j+1)<\sigma_i\tilde\sigma(j). \label{cond:FR-F}
\end{eqnarray}
Comparing (\ref{cond:FR-F}) with (\ref{cond:F}), we find that the
only difference between them is the transposition $\sigma_i$
factor in the ``if" conditions.  For a given $\tilde\sigma\in
{\cal S}_N$ with $\tilde\sigma(j)=i$ and $\tilde\sigma(k)=i+1$,
one finds that if $|j-k|>1$, then $\sigma_i$ does not affect the
sequence of $\alpha_{\tilde\sigma}$ at all, that is, the sign of
inequality $``>"$ or ``$\geq$" between two neighboring root
indexes is unchanged with the action of $\sigma_i$; and if
$|j-k|=1$, then in the summation sequences of
$\alpha_{\tilde\sigma}$, when $\tilde\sigma(j+1)=i+1$ and
$\tilde\sigma(j)=i$, sign ``$\geq$" changes to $``>"$, while when
$\tilde\sigma(j+1)=i$ and $\tilde\sigma(j)=i+1$, $``>"$ changes to
``$\geq$". Thus (\ref{cond:F}) and (\ref{eq:FR-F}) differ only
when equal labels $\alpha_{\tilde\sigma}$ appear. With the help of
the $c$-number $S$ in the definition of $F_{1,\ldots, N}$
(\ref{de:F}) and the relation $c_{12}c_{21}=1$, one easily shows
that for the equal labels, $F_{\sigma_i(1\ldots
N)}R^{\sigma_i}_{1\ldots N}- F_{1\ldots N}=0$. (For detailed
proof, please see \cite{zsy04}.)

Therefore, we have proved that the Drinfeld twist factorizes the
$R$-matrix of the $N$-site $gl(m|n)$ model:
\begin{eqnarray}
R_{1\ldots N}^\sigma(z_1,\ldots,z_N)
 =F^{-1}_{\sigma(1\ldots N)}(z_{\sigma(1)},\ldots,
   z_{\sigma(N)})F_{1\ldots N}(z_1,\ldots,z_N).
\end{eqnarray}
In summary, the factorizing $F$-matrix $F_{1\ldots N}$ of the
$gl(m|n)$ model is proved to satisfy all three properties.%\\[3mm]

\subsection{Inverse $F^{-1}_{1\ldots N}$ of the $F$-matrix }

The non-degenerate property of the $F$-matrix implies that we can
find the inverse matrix $F^{-1}_{1\ldots N}$. To do so, we first
define
\begin{eqnarray}
F^*_{1\ldots N}&=&\sum_{\sigma\in {\cal S}_N}
   \sum_{\alpha_{\sigma(1)}\ldots\alpha_{\sigma(N)}}^{\quad\quad **}
   S(c,\sigma,\alpha_\sigma)R_{\sigma(1\ldots N)}^{\sigma^{-1}}
   \prod_{j=1}^N P_{\sigma(j)}^{\alpha_{\sigma(j)}},
    \label{de:F*}  \nonumber\\
\end{eqnarray}
where the sum $\sum^{**}$ is taken over all possible $\alpha_i$
which satisfies the following non-increasing constraints:
\begin{eqnarray}
&& \alpha_{\sigma(i+1)}\leq \alpha_{\sigma(i)}\quad \mbox{if}\quad
              \sigma(i+1)<\sigma(i), \nonumber\\
&& \alpha_{\sigma(i+1)}< \alpha_{\sigma(i)}\quad \mbox{if}\quad
              \sigma(i+1)>\sigma(i). \label{cond:F*}
\end{eqnarray}
%and $S^*(c,\sigma,\alpha_\sigma)$ is given by
%\begin{eqnarray}
%S^*(c,\sigma,\alpha_\sigma)=\exp\{\sum_{k>l=1}^N\delta^{[3]}_{\alpha_{\sigma(k)},\alpha_{\sigma(l)}}
%    \ln(1+c_{\sigma(k)\sigma(l)})\}
%\end{eqnarray}

Now we compute the product of $F_{1\ldots N}$ and $F^*_{1\ldots
N}$. Substituting (\ref{de:F}) and (\ref{de:F*}) into the product,
we have
\begin{eqnarray}
 F_{1\ldots N}F^*_{1\ldots N}% \nonumber\\
 &=&\sum_{\sigma\in {\cal S}_N}\sum_{\sigma'\in {\cal S}_N}
    \sum^{\quad\quad *}_{\alpha_{\sigma_1}\ldots\alpha_{\sigma_N}}
    \sum^{\quad\quad **}_{\beta_{\sigma'_1}\ldots\beta_{\sigma'_{N}}}
    S(c,\sigma,\alpha_\sigma)S(c,\sigma',\beta_{\sigma'})
    \nonumber\\ &&\times
    \prod_{i=1}^N P_{\sigma(i)}^{\alpha_{\sigma(i)}}
    R^{\sigma}_{1\ldots N}R^{{\sigma'}^{-1}}_{\sigma'(1\ldots N)}
    \prod_{i=1}^N P_{\sigma'(i)}^{\beta_{\sigma'(i)}} \nonumber\\
 &=&\sum_{\sigma\in {\cal S}_N}\sum_{\sigma'\in {\cal S}_N}
    \sum^{\quad\quad *}_{\alpha_{\sigma_1}\ldots\alpha_{\sigma_N}}
    \sum^{\quad\quad **}_{\beta_{\sigma'_1}\ldots\beta_{\sigma'_{N}}}
    S(c,\sigma,\alpha_\sigma)S(c,\sigma',\beta_{\sigma'})
     \nonumber\\ &&\times
    \prod_{i=1}^N P_{\sigma(i)}^{\alpha_{\sigma(i)}}
    R^{{\sigma'}^{-1}\sigma}_{\sigma'(1\ldots N)}
    \prod_{i=1}^N P_{\sigma'(i)}^{\beta_{\sigma'(i)}}. \label{eq:FF*-1}
\end{eqnarray}
To evaluate the r.h.s., we examine the matrix element of the
$R$-matrix
\begin{eqnarray}
\left(R^{{\sigma'}^{-1}\sigma}_{\sigma'(1\ldots N)}\right)
  ^{\alpha_{\sigma(N)}\ldots\alpha_{\sigma(1)}}
  _{\beta_{\sigma'(N)}\ldots\beta_{\sigma'(1)}}.
  \label{eq:R-index}
\end{eqnarray}
Note that the sequence $\{ \alpha_{\sigma}\}$ is non-decreasing
and $\{\beta_{\sigma'}\}$ is non-increasing. Thus the
non-vanishing condition of the matrix element (\ref{eq:R-index})
requires that $\alpha_{\sigma}$ and $\beta_{\sigma'}$ satisfy
\begin{eqnarray}
\beta_{\sigma'(N)}=\alpha_{\sigma(1)},\ldots,
\beta_{\sigma'(1)}=\alpha_{\sigma(N)}. \label{re:alpha-beta}
\end{eqnarray}
One can verify \cite{Albert00} that (\ref{re:alpha-beta}) is
fulfilled only if
\begin{eqnarray}
\sigma'(N)=\sigma(1),\ldots,\sigma'(1)=\sigma(N).
\label{re:sigma-sigma'}
\end{eqnarray}

Let $\bar\sigma$ be the maximal element of the ${\cal S}_N$ which
reverses the site labels
\begin{eqnarray}
\bar\sigma(1,\ldots,N)=(N,\ldots,1).
\end{eqnarray}
Then from (\ref{re:sigma-sigma'}), we have
\begin{eqnarray}
\sigma'=\sigma\bar\sigma. \label{eq:sigma'}
\end{eqnarray}
Substituting (\ref{re:alpha-beta}) and (\ref{eq:sigma'}) into
(\ref{eq:FF*-1}), we have
\begin{eqnarray}
 F_{1\ldots N}F^*_{1\ldots N}%= \nonumber\\
 &=&\sum_{\sigma\in {\cal S}_N}
    \sum^{\quad\quad *}_{\alpha_{\sigma_1}\ldots\alpha_{\sigma_N}}
    S(c,\sigma,\alpha_\sigma)S(c,\sigma,\alpha_\sigma)
    \prod_{i=1}^N P_{\sigma(i)}^{\alpha_{\sigma(i)}}
    R^{\bar\sigma}_{\sigma(N\ldots 1)}
    \prod_{i=1}^N P_{\sigma(i)}^{\alpha_{\sigma(i)}}\label{eq:FF*-2}.
      \nonumber\\
\end{eqnarray}
The decomposition of $R^{\bar\sigma}$ in terms of elementary
$R$-matrices is unique module GYBE. One reduces from
(\ref{eq:FF*-2}) that $FF^*$ is a diagonal matrix:
\begin{eqnarray}
F_{1\ldots N}F^*_{1\ldots N}=\prod_{i<j}\Delta_{ij},
\end{eqnarray}
where
\begin{eqnarray}
[\Delta_{ij}]^{\beta_i\beta_j}_{\alpha_i\alpha_j} =
 \delta_{\alpha_i\beta_i}\delta_{\alpha_j\beta_j}\left\{
 \begin{array}{cl}
  a_{ij}& \mbox{if} \ \alpha_i>\alpha_j\\
  a_{ji}& \mbox{if} \ \alpha_i<\alpha_j\\
  4a_{ij}a_{ji}& \mbox{if}\ \alpha_i=\alpha_j=1,2,\ldots,m\\
  1& \mbox{if}\ \alpha_i=\alpha_j=m+1,\ldots,m+n
  \end{array}\right.. \label{de:F-Inv}
\end{eqnarray}

Therefore, the inverse of the $F$-matrix is given by
\begin{equation}
F^{-1}_{1\ldots N}=F^*_{1\ldots N}\prod_{i<j}\Delta_{ij}^{-1}.
\end{equation}

\sect{The monodromy matrix in the $F$-basis} ~~~ In the previous
section, we see that the $gl(m|n)$ $R$-matrix factorizes in terms
of the $F$-matrix and its inverse which we constructed explicitly.
The column vectors of the inverse of the $F$-matrix form a set of
basis on which $gl(m|n)$ acts. In this section, we study the
generators of $gl(m|n)$ and the elements of the monodromy matrix
in the $F$-basis.

\subsection{$gl(m|n)$ generators in the $F$-basis}
Denoted by $E^{\gamma\ \gamma\pm1}$ the simple generators of the
$N$-site $gl(m|n)$ supersymmetric system. Then $E^{\gamma\
\gamma\pm 1}=E^{\gamma\,\gamma\pm 1}_{(1)} +\ldots+E^{\gamma\
\gamma\pm 1}_{(N)}$, where $E^{\gamma\,\gamma\pm 1}_{(k)}$ acts on
the $k$th component of the tensor product space. Let $\tilde
E^{\gamma\,\gamma\pm 1}$ denote the corresponding simple
generators in the $F$-basis: $\tilde E^{\gamma\, \gamma\pm
1}=F_{1\ldots N}E^{\gamma\, \gamma\pm 1}F_{1\ldots N}^{-1}$. For
later use, we derive $\tilde E^{\gamma\ \gamma+1}$. From the
expressions of $F$ and its inverse, we have
\begin{eqnarray}
 \tilde E^{\gamma\ \gamma+1}%\nonumber\\
 &=&F_{1\ldots N}E^{\gamma\ \gamma+1}F^{-1}_{1\ldots N}\nonumber\\
 &=&\sum_{\sigma,\sigma'\in {\cal S}_N}
    \sum^{\quad\quad*}_{\alpha_{\sigma(1)}\ldots\alpha_{\sigma(N)}}
    \sum^{\quad\quad**}_{\beta_{\sigma'(1)}\ldots\beta_{\sigma'(N)}}
    S(c,\sigma,\alpha_\sigma)S(c,\sigma',\beta_{\sigma'})
    \nonumber\\ && \times
    \prod_{i=1}^N P_{\sigma(i)}^{\alpha_{\sigma(i)}}
    R^{\sigma}_{1\ldots N}E^{\gamma\ \gamma+1}
    R^{{\sigma'}^{-1}}_{\sigma'(1\ldots N)}
    \prod_{i=1}^N
    P_{\sigma'(i)}^{\beta_{\sigma'(i)}}\prod_{i<j}\Delta_{ij}^{-1}
     \nonumber\\
 &=&\sum_{\sigma,\sigma'\in {\cal S}_N}
    \sum^{\quad\quad*}_{\alpha_{\sigma(1)}\ldots\alpha_{\sigma(N)}}
    \sum^{\quad\quad**}_{\beta_{\sigma'(1)}\ldots\beta_{\sigma'(N)}}
    S(c,\sigma,\alpha_\sigma)S(c,\sigma',\beta_{\sigma'})
    \nonumber\\ &&\times
    \prod_{i=1}^N P_{\sigma(i)}^{\alpha_{\sigma(i)}}E^{\gamma\ \gamma+1}
    R^{{\sigma'}^{-1}\sigma}_{\sigma'(1\ldots N)}
    \prod_{i=1}^N P_{\sigma'(i)}^{\beta_{\sigma'(i)}}
    \prod_{i<j}\Delta_{ij}^{-1}
    \label{eq:E-1}     \\
 &=&\sum_{\sigma,\sigma'\in {\cal S}_N}
    \sum_{k=1}^N E^{\gamma\ \gamma+1}_{(\sigma(a+l))}
    \sum^{\quad\quad*}_{\alpha_{\sigma(1)}\ldots\alpha_{\sigma(N)}}
    \sum^{\quad\quad**}_{\beta_{\sigma'(1)}\ldots\beta_{\sigma'(N)}}
    S(c,\sigma,\alpha_\sigma)S(c,\sigma',\beta_{\sigma'})
     \nonumber\\ && \times
    P_{\sigma(1)}^{\alpha_{\sigma(1)}}
    P_{\sigma(2)}^{\alpha_{\sigma(2)}}\ldots
    P_{\sigma(a)}^{\alpha_{\sigma(a)}=\gamma-1}
    P_{\sigma(a+1)}^{\alpha_{\sigma(a+1)}=\gamma}\ldots
    \left(P_{\sigma(a+l)=k}^{\alpha_{\sigma(a+l)}
           =\gamma\rightarrow \gamma+1}\right)\ldots
    \nonumber\\&& \times
     P_{\sigma(N)}^{\alpha_{\sigma(N)}}
    R^{{\sigma'}^{-1}\sigma}_{\sigma'(1\ldots N)}
    \prod_{i=1}^N
    P_{\sigma'(i)}^{\beta_{\sigma'(i)}}\prod_{i<j}\Delta_{ij}^{-1},
    \label{eq:E-2} \nonumber\\
\end{eqnarray}
where in (\ref{eq:E-1}), we have used $[E^{\gamma\,\gamma\pm
1},R^{\sigma}_{1\ldots N}]=0.$ The element of
$R^{{\sigma'}^{-1}\sigma}_{\sigma'(1\ldots N)}$ between $
P_{\sigma(1)}^{\alpha_{\sigma(1)}}\ldots
    \left(P_{\sigma(a+l)=k}^{\alpha_{\sigma(a+l)}
          =\gamma\rightarrow\gamma+1}\right)
    \ldots P_{\sigma(N)}^{\alpha_{\sigma(N)}}$ and $
 P_{\sigma'(N)}^{\beta_{\sigma'(N)}}\ldots
   P_{\sigma'(1)}^{\beta_{\sigma'(1)}}$
is denoted as
 \begin{eqnarray}
 \left(R^{{\sigma'}^{-1}\sigma}_{\sigma'(1\ldots N)}\right)
 ^{\stackrel{\sigma(N)}{\alpha_{\sigma(N)}}\ldots
 \stackrel{\sigma(a+l)=k}{\gamma\rightarrow\gamma+1}\ldots
 \stackrel{\sigma(a+1)}{\gamma}
 \stackrel{\sigma(a)}{\gamma-1}\ldots
  \stackrel{\sigma(1)}{\alpha_{\sigma(1)}}}
 _{\beta_{\sigma'(N)}\ldots \beta_{\sigma'(1)}}. \label{eq:E-R-index}
 \end{eqnarray}
We call the sequence $\{\alpha_{\sigma(l)}\}$ {\bf normal} if it
is arranged according to the rules in (\ref{cond:F}), otherwise,
we call it {\bf abnormal}.

It is now convenient for us to discuss the non-vanishing condition
of the $R$-matrix element (\ref{eq:E-R-index}).  Comparing
(\ref{eq:E-R-index}) with (\ref{eq:R-index}), we find that the
difference between them lies in the $k$th site. Because the group
label in the $k$th space has been changed, the sequence
$\{\alpha_\sigma\}$ is now a abnormal sequence. However, it can be
permuted to the normal sequence by some permutation $\hat
\sigma_k$. Namely, $\alpha_{\gamma\rightarrow\gamma+1}$ in the
abnormal sequence  can be moved to a suitable position by using
the permutation $\hat \sigma_k$ according to rules in
(\ref{cond:F}). (It is easy to verify that $\hat\sigma_k$ is
unique by using (\ref{cond:F}).) Thus, by procedure similar to
that in the previous section, we find that when
\begin{equation}
\sigma'=\hat\sigma_k\sigma\bar\sigma\quad \mbox{and}\quad
 \beta_{\sigma'(N)}=\alpha_{\sigma(1)},\ldots,
 \beta_{\sigma'(1)}=\alpha_{\sigma(N)},\label{eq:E-sigma}
\end{equation}
the $R$-matrix element (\ref{eq:E-R-index}) is non-vanishing.
% (We note that for
%a $\{P_{\sigma}^{\alpha_\sigma}\}$ sequence
%(\ref{eq:sqc-alpha}), we can find $m-l+1$
%$\{P_{\sigma'}^{\beta_\sigma'}\}$ sequences which ensure the
%non-vanishing of $R$-matrix according to the (\ref{cond:F}) and
%(\ref{cond:F*})).

Because the non-zero condition of the elementary $R$-matrix
element $R^{i'j'}_{ij}$ is $i+j=i'+j'$, the following $R$-matrix
elements
\begin{eqnarray}
 \left(R^{{\sigma'}^{-1}\sigma}_{\sigma'(1\ldots N)}\right)
 ^{\stackrel{\sigma(N)}{\alpha_{\sigma(N)}}\ldots
 \stackrel{\sigma(a+l)=k}{\gamma} \ldots
 \stackrel{\sigma(n)}{\gamma\rightarrow\gamma+1}\ldots
 \stackrel{\sigma(a+1)}{\gamma}
 \stackrel{\sigma(a)}{\gamma-1}\ldots
  \ldots \stackrel{\sigma(1)}{\alpha_{\sigma(1)}}}
 _{\beta_{\sigma'(N)}\ldots \beta_{\sigma'(1)}} \label{eq:E-R-nindex}
 \end{eqnarray}
with $a+1\le n<a+l$ are also non-vanishing.

Therefore, (\ref{eq:E-2}) becomes
\begin{eqnarray}
 \tilde E^{\gamma\ \gamma+1}%\nonumber\\
 &=&\sum_{\sigma\in {\cal S}_N}\sum_{k=1}^N
    \sum^{\quad\quad *}_{\alpha_{\sigma_1}\ldots\alpha_{\sigma_N}}
    S(c,\sigma,\alpha_\sigma)S(c,\hat\sigma_k\sigma,\alpha_{\hat\sigma_k\sigma})
       \nonumber\\ && \times
    \left[E^{\gamma\ \gamma+1}_{(\sigma(a+l))}
    P_{\sigma(1)}^{\alpha_{\sigma(1)}}\ldots
    P_{\sigma(a)}^{\gamma-1}
    P_{\sigma(a+1)}^{\gamma}\ldots
    P_{\sigma(a+l)=k}^{\gamma\rightarrow\gamma+1}
    \ldots P_{\sigma(N)}^{\alpha_{\sigma(N)}}+\ldots\right.
       \nonumber\\ && \mbox{}
   +E^{\gamma\ \gamma+1}_{(\sigma(a+n))}
    P_{\sigma(1)}^{\alpha_{\sigma(1)}}\ldots
    P_{\sigma(a)}^{\gamma-1}
    P_{\sigma(a+1)}^{\gamma}\ldots
    P_{\sigma(a+n)}^{\gamma\rightarrow\gamma+1}
    \ldots P_{\sigma(a+l)=k}^{\gamma}\ldots
    P_{\sigma(N)}^{\alpha_{\sigma(N)}}+ \ldots
       \nonumber\\ && \mbox{}  \mbox{}\left.
   +E^{\gamma\ \gamma+1}_{(\sigma(a+1))}
    P_{\sigma(1)}^{\alpha_{\sigma(1)}}\ldots
    P_{\sigma(a)}^{\gamma-1}
    P_{\sigma(a+1)}^{\gamma\rightarrow\gamma+1}
    \ldots P_{\sigma(a+l)=k}^{\gamma}\ldots
    P_{\sigma(N)}^{\alpha_{\sigma(N)}}\right]
        \nonumber\\ && \times
    R^{\bar\sigma\sigma^{-1}\hat\sigma^{-1}_k\sigma}
     _{\hat\sigma_k\sigma(N\ldots 1)}
    \prod_{i=1}^N
    P_{\hat\sigma_k\sigma(i)}^{\alpha_{\hat\sigma_k\sigma(i)}}
    \prod_{i<j}\Delta_{ij}^{-1} \label{eq:E-3}\\
 &=&\sum_{k=1}^N E^{\gamma\ \gamma+1}_{(k)}\otimes_{j\ne k}
   G^{\gamma\ \gamma+1}_{(j)}(k,j).
     \label{eq:E-12-tilde}
\end{eqnarray}
Here in (\ref{eq:E-3}) $\hat\sigma_k$ is the element of  ${\cal
S}_N$ which permutes the first abnormal sequence in the square
bracket of (\ref{eq:E-3}) to normal sequence and $G^{\gamma\
\gamma+1}(i,j)$ in (\ref{eq:E-12-tilde}) has the following
elements: For $1<\gamma+1\leq m$,
\begin{equation}
  G^{\gamma\ \gamma+1}(i,j)_{kl}
  =\delta_{k,l}\left\{\begin{array}{cl}
   2& k=\gamma\\
   (2a_{ij})^{-1}&  k=\gamma+1\\
   1& \mbox{ otherwise }\end{array}\right. ,
\end{equation}
for $\gamma=m$
\begin{equation}
  G^{\gamma\ \gamma+1}(i,j)_{kl}
  =\delta_{k,l}\left\{\begin{array}{cl}
   2&  k=\gamma\\
   1& \mbox{ otherwise }\end{array}\right. ,
\end{equation}
and for $m+1\leq\gamma< m+n $
\begin{equation}
  G^{\gamma\ \gamma+1}(i,j)_{kl}
  =\delta_{k,l}\left\{\begin{array}{cl}
   (a_{ij})^{-1}& k=\gamma\\
   1& \mbox{ otherwise }\end{array}\right. .
\end{equation}

The non-simple generators are generated by the simple generators
through the relation
\begin{eqnarray}
 \tilde E^{m+n-\alpha\ m+n}
 &=&[\tilde E^{m+n-\alpha\ m+n-\alpha+1},
  [\tilde E^{m+n-\alpha+1\ m+n-\alpha+2},[\ldots,\nonumber\\
&&[\tilde E^{m+n-2\ m+n-1},\tilde E^{m+n-1\ m+n}]\ldots]]].
\end{eqnarray}
We have:\\
1. For $m=0$, the $gl(m|n)$ supersymmetric model degenerates to
the bosonic $gl(n)$ model, which has been discussed in
\cite{Albert00} by Albert et al.\\
2. For $n=0$, $\alpha<m$,
\begin{eqnarray}
\tilde E^{m-\alpha\ m}&=&
 \sum_{k=1}^{\alpha}\sum_{i_1\ne\ldots\ne i_k}\prod_{\gamma=1}^{k-1}
 {\eta\over z_{i_{\gamma}}-z_{i_{\gamma+1}}}
 \sum_{\alpha=\beta_0>\ldots>\beta_k=0}
 \otimes_{l=1}^k E^{m-\beta_{l-1}\ m-\beta_{l}}_{(i_l)}
  \nonumber\\ && \otimes \mbox{diag}\left(
 1,\ldots,1,2,\underbrace{a_{i_1j}^{-1},\ldots,a_{i_1j}^{-1}}_
 {\beta_0-\beta_1},\ldots,
 \underbrace{a_{i_{k-1}j}^{-1},\ldots,a_{i_{k-1}j}^{-1}}_
 {\beta_{k}-\beta_{k-1}},\right.
 \nonumber\\ &&\quad\quad\quad\left.
 \underbrace{a_{i_kj}^{-1},\ldots,a_{i_kj}^{-1}}_
 {\beta_{k-1}-\beta_k-1},
 (2a_{i_kj})^{-1}\right)_{(j)},
% \nonumber\\
\end{eqnarray}
3. For $m,n\geq 1$ and $\alpha<n$,
\begin{eqnarray}
\tilde E^{m+n-\alpha\ m+n}&=&
 \sum_{k=1}^{\alpha}\sum_{i_1\ne\ldots\ne i_k}\prod_{\gamma=1}^{k-1}
 {\eta\over z_{i_{\gamma+1}}-z_{i_\gamma}}
 \sum_{\alpha=\beta_0>\ldots>\beta_k=0}
 \otimes_{l=1}^k E^{m+n-\beta_{l-1}\ m+n-\beta_{l}}_{(i_l)}
  \nonumber\\ && \otimes \mbox{diag}\left(
 1,\ldots,1,\underbrace{a_{i_1j}^{-1},\ldots,a_{i_1j}^{-1}}_
 {\beta_0-\beta_1},\ldots,
 \underbrace{a_{i_kj}^{-1},\ldots,a_{i_kj}^{-1}}_
 {\beta_{k-1}-\beta_k},1\right)_{(j)},
\end{eqnarray}
%for $\alpha=n$,
%\begin{eqnarray}
%\tilde E^{m+n-\alpha\ m+n}&=&
% \sum_{k=1}^{\alpha}\sum_{i_1\ne\ldots\ne i_k}\prod_{\gamma=1}^{k-1}
% {\eta\over z_{i_{\gamma+1}}-z_{i_\gamma}}
% \sum_{\alpha=\beta_0>\ldots>\beta_k=0}
 %\otimes_{l=1}^k E^{m+n-\beta_{l-1}\ m+n-\beta_{l}}_{(i_l)}
 %% \nonumber\\ && \otimes \mbox{diag}(
 %1,\ldots,1,2,\underbrace{a_{i_1j}^{-1},\ldots,a_{i_1j}^{-1}}_
 %{\beta_0-\beta_1-1},
 %\underbrace{a_{i_2j}^{-1},\ldots,a_{i_2j}^{-1}}_
 %{\beta_1-\beta_2},
 %\ldots,
 %\underbrace{a_{i_kj}^{-1},\ldots,a_{i_kj}^{-1}}_
 %{\beta_{k-1}-\beta_k},1), \nonumber\\
%\end{eqnarray}
4. For $m,n\geq 1$ and  $\alpha\geq n$,
\begin{eqnarray}
\tilde E^{m+n-\alpha\ m+n}&=&
 \sum_{k=1}^{\alpha}\sum_{i_1\ne\ldots\ne i_k}
 \prod_{\gamma=1}^{\mbox{min}(k-1,n-1)}
 {\eta\over z_{i_{\gamma+1}}-z_{i_\gamma}}
 \prod_{\gamma=n+1}^{k-1}
 {\eta\over z_{i_\gamma}-z_{i_{\gamma+1}}}
 \nonumber\\ && \times
 \sum_{\alpha=\beta_0>\ldots>\beta_k=0}
 \otimes_{l=1}^k E^{m+n-\beta_{l-1}\ m+n-\beta_{l}}_{(i_l)}
  \nonumber\\ && \otimes \mbox{diag}\left(
 1,\ldots,1,2,
 \underbrace{a_{i_1j}^{-1},\ldots,a_{i_1j}^{-1}}_
 {\beta_0-\beta_1},\ldots,
 \underbrace{a_{i_pj}^{-1},\ldots,a_{i_pj}^{-1}}_
 {\beta_{p-1}-\beta_p},\right.
  \nonumber\\ && \quad\quad \left.
 \underbrace{a_{i_{p+1}j}^{-1},\ldots,a_{i_{p+1}j}^{-1}}_
 {\begin{array}{c}\beta_{p}-\beta_{p+1}-1\\
    \beta_{p}\geq n\mbox{ and }\beta_{p+1}<n \end{array}},
 \underbrace{a_{i_{p+2}j}^{-1},\ldots,a_{i_{p+2}j}^{-1}}_
 {\beta_{p+1}-\beta_{p+2}},\ldots,
 \underbrace{a_{i_kj}^{-1},\ldots,a_{i_kj}^{-1}}_
 {\beta_{k-1}-\beta_k},1\right)_{(j)}.\nonumber\\
\end{eqnarray}

\subsection{Elements of the monodromy matrix in the $F$-basis}

In the $F$-basis, the monodromy matrix $T(u)$, (\ref{de:T-marix}),
becomes
\begin{eqnarray}
 \tilde {T}(u)=\left(\begin{array}{cccc}
\tilde { A}_{11}(u)& \cdots& \tilde {A}_{1\ m+n-1}(u)&
    \tilde { B}_1(u)\\
 \vdots & \vdots & \vdots&\vdots\\
\tilde { A}_{m+n-1\ 1}(u)& \cdots& \tilde { A}_{m+n-1\ m+n-1}(u)&
    \tilde { B}_{m+n-1}(u)\\
\tilde { C}_{1}(u)& \cdots&\tilde { C}_{m+n-1}(u)&
    \tilde { D}(u)
 \end{array}\right). \label{de:T-marix-tilde}
 \end{eqnarray}
We first study the diagonal element $\tilde D(u)$. Acting the
$F$-matrix on $D(u)$, we have
\begin{eqnarray}
 F_{1\ldots N}D
 &=&\sum_{\sigma\in {\cal S}_N}
    \sum_{\alpha_{\sigma(1)}\ldots\alpha_{\sigma(N)}}^{\quad\quad*}
    S(c,\sigma,\alpha_\sigma)\prod_{i=1}^N
    P_{\sigma(i)}^{\alpha_{\sigma}}R^\sigma_{1\ldots N}
    P_0^{m+n} T_{0,1\ldots N}P_0^{m+n} \nonumber\\
 &=&\sum_{\sigma\in {\cal S}_N}
    \sum_{\alpha_{\sigma(1)}\ldots\alpha_{\sigma(N)}}^{\quad\quad*}
    S(c,\sigma,\alpha_\sigma)\prod_{i=1}^N
    P_{\sigma(i)}^{\alpha_{\sigma}}
    P_0^{m+n} T_{0,\sigma(1\ldots N)}P_0^{m+n}
    R^\sigma_{1\ldots N}.
\end{eqnarray}
Following \cite{Albert00}, we can split the sum $\sum^*$ according
to the number of occurrences of the index $m+n$.
\begin{eqnarray}
F_{1\ldots N}T^{{m+n}\ {m+n}}
 &=&\sum_{\sigma\in {\cal S}_N}
    \sum_{k=0}^N
    \sum_{\alpha_{\sigma(1)}\ldots\alpha_{\sigma(N)}}^{\quad\quad*}
    S(c,\sigma,\alpha_\sigma)
    \prod_{j=N-k+1}^N \delta_{\alpha_{\sigma(j)},{m+n}}
    P_{\sigma(j)}^{\alpha_{\sigma(j)}} \nonumber\\ &&\times
    \prod_{j=1}^{N-k}P_{\sigma(j)}^{\alpha_{\sigma(j)}}
    P_0^{m+n} T_{0,\sigma(1\ldots N)}P_0^{m+n}
    R^\sigma_{1\ldots N}. \label{eq:T-tilde-1}
\end{eqnarray}
Consider the prefactor of $R^\sigma_{1\ldots N}$. We have
\begin{eqnarray}
 &&\prod_{j=1}^{N-k}P_{\sigma(j)}^{\alpha_{\sigma(j)}}
   \prod_{j=N-k+1}^N
    P_{\sigma(j)}^{{m+n}}
    P_0^{m+n} T_{0,\sigma(1\ldots N)}P_0^{m+n}\nonumber\\
 &=&\prod_{j=1}^{N-k}P_{\sigma(j)}^{\alpha_{\sigma(j)}}
   \prod_{j=N-k+1}^N\left(
   R_{0\,\sigma(j)}\right)^{{m+n}\ {m+n}}_{{m+n}\ {m+n}}
   P_0^{m+n} T_{0,\sigma(1\ldots N-k)}P_0^{m+n}
   \prod_{j=N-k+1}^N P_{\sigma(j)}^{{m+n}} \nonumber\\
 &=&
   \prod_{j=1}^{N-k}P_{\sigma(j)}^{\alpha_{\sigma(j)}}
   P_0^{m+n} T_{0,\sigma(1\ldots N-k)}P_0^{m+n}
   \prod_{j=N-k+1}^N P_{\sigma(j)}^{{m+n}} \nonumber\\
 &=&
   \prod_{i=1}^{N-k}\left(R_{0\,\sigma(i)}\right)
            ^{{m+n}\alpha_{\sigma(i)}}_{{m+n}\alpha_{\sigma(i)}}
   \prod_{j=1}^{N-k}P_{\sigma(j)}^{\alpha_{\sigma(j)}}
   \prod_{j=N-k+1}^N P_{\sigma(j)}^{{m+n}}\nonumber\\
 &=&
   \prod_{i=1}^{N-k}a_{0\,\sigma(i)}
   \prod_{j=1}^{N-k}P_{\sigma(j)}^{\alpha_{\sigma(j)}}
   \prod_{j=N-k+1}^N P_{\sigma(j)}^{{m+n}},  \label{eq:T-tilde-2}
\end{eqnarray}
where $c_{0i}=c(u,z_i)$, $a_{0i}=a(u,z_i)$. Substituting
(\ref{eq:T-tilde-2}) into (\ref{eq:T-tilde-1}), we have
\begin{eqnarray}
F_{1\ldots N}T^{{m+n}{m+n}}=\otimes_{i=1}^N
  \mbox{diag}\left(a_{0i},\ldots,a_{0i},1\right)_{(i)}
 F_{1\ldots N}.
\end{eqnarray}
Therefore,
\begin{eqnarray}
\tilde D(u)=\otimes_{i=1}^N
  \mbox{diag}\left(a_{0i},\ldots,a_{0i},1\right)_{(i)}.
  \label{eq:T33-tilde}
\end{eqnarray}

The creation operators in the monodromy matrix can then be
obtained as follows:
\begin{equation}
\tilde C_{\alpha}(u)=[\tilde E^{\alpha\,{m+n}},\tilde D(u)],\quad
(1\leq \alpha<m+n),
\end{equation}
which follows from the $gl(m|n)$ invariance of the $R$-matrix,
i.e. in terms of the monodromy matrix,
\begin{equation}
[\tilde T(u), \tilde E^{\alpha \beta}_{(0)}+\tilde E^{\alpha
\beta}]=0.
\end{equation}
 Substituting $\tilde
E^{\alpha\,{m+n}}$, $\tilde E^{{m+n}\,\alpha}$ and $\tilde
T^{{m+n}\ {m+n}}$ into the above relations yields\\
for $n=0$,
\begin{eqnarray}
\tilde C_{m-\alpha}(u) &=&
 -\sum_{k=1}^{\alpha}\sum_{i_1\ne\ldots\ne i_k}b_{0i_{k}}
 \prod_{\gamma=1}^{k-1}
 {a_{0i_{\gamma}}\eta\over z_{i_{\gamma}}-z_{i_\gamma+1}}
 \sum_{\alpha=\beta_0>\ldots>\beta_k=0}
 \otimes_{l=1}^k E^{m-\beta_{l-1}\ m-\beta_{l}}_{(i_l)}
  \nonumber\\ && \otimes \mbox{diag}\left(
 a_{0j},\ldots,a_{0j},2a_{0j},
 \underbrace{a_{0j}a_{i_1j}^{-1},\ldots,a_{0j}a_{i_1j}^{-1}}_
 {\beta_0-\beta_1},\right.
 \nonumber\\ && \quad\quad \ldots, \left.
 \underbrace{a_{0j}a_{i_kj}^{-1},\ldots,a_{0j}a_{i_kj}^{-1}}_
 {\beta_{k-1}-\beta_k-1},c_{0j}(2a_{i_k,j})^{-1}\right)_{(j)},
 \nonumber\\
 \label{eq:C-alpha-Fermin}
\end{eqnarray}
for $m,n\geq 1$ and $\alpha<n$,
\begin{eqnarray}
\tilde C_{m+n-\alpha}(u)&=&
 \sum_{k=1}^{\alpha}\sum_{i_1\ne\ldots\ne i_k}b_{0i_{k}}
 \prod_{\gamma=1}^{k-1}
 {a_{0i_{\gamma}}\eta\over z_{i_{\gamma+1}}-z_{i_\gamma}}
 \sum_{\alpha=\beta_0>\ldots>\beta_k=0}
 \otimes_{l=1}^k E^{m+n-\beta_{l-1}\ m+n-\beta_{l}}_{(i_l)}
  \nonumber\\ && \otimes \mbox{diag}\left(
 a_{0j},\ldots,a_{0j},
 \underbrace{a_{0j}a_{i_1j}^{-1},\ldots,a_{0j}a_{i_1j}^{-1}}_
 {\beta_0-\beta_1},\ldots,
 \underbrace{a_{0j}a_{i_kj}^{-1},\ldots,a_{0j}a_{i_kj}^{-1}}_
 {\beta_{k-1}-\beta_k},1\right)_{(j)},\nonumber\\
 \label{eq:C-alpha-Bose}
\end{eqnarray}
and for $m,n\geq 1$ and $\alpha\geq n$,
\begin{eqnarray}
\tilde C_{m+n-\alpha}(u)&=&
 \sum_{k=1}^{\alpha}\sum_{i_1\ne\ldots\ne i_k}b_{0i_k}
 \prod_{\gamma=1}^{\mbox{min}(k-1,n-1)}
 {a_{0i_\gamma}\eta\over z_{i_{\gamma+1}}-z_{i_\gamma}}
 \prod_{\gamma=n+1}^{k-1}
 {a_{0i_\gamma}\eta\over z_{i_\gamma}-z_{i_{\gamma+1}}}
 \nonumber\\ && \times
 \sum_{\alpha=\beta_0>\ldots>\beta_k=0}
 \otimes_{l=1}^k E^{m+n-\beta_{l-1}\ m+n-\beta_{l}}_{(i_l)}
  \nonumber\\ && \otimes \mbox{diag}\left(
 a_{0j},\ldots,a_{0j},2a_{0j},
 \underbrace{a_{0j}a_{i_1j}^{-1},\ldots,a_{0j}a_{i_1j}^{-1}}_
 {\beta_0-\beta_1},\ldots,
 \underbrace{a_{0j}a_{i_pj}^{-1},\ldots,a_{0j}a_{i_pj}^{-1}}_
 {\beta_{p-1}-\beta_p},\right. \nonumber\\ && \quad\quad
 \underbrace{a_{0j}a_{i_{p+1}j}^{-1},\ldots,a_{0j}a_{i_{p+1}j}^{-1}}_
 {\begin{array}{c}\beta_{p}-\beta_{p+1}-1\\
    \beta_{p}\geq n\mbox{ and }\beta_{p+1}<n \end{array}},
 \underbrace{a_{0j}a_{i_{p+2}j}^{-1},\ldots,a_{0j}a_{i_{p+2}j}^{-1}}_
 {\beta_{p+1}-\beta_{p+2}},
 \nonumber\\ && \quad\quad\ldots, \left.
 \underbrace{a_{0j}a_{i_kj}^{-1},\ldots,a_{0j}a_{i_kj}^{-1}}_
 {\beta_{k-1}-\beta_k},1\right)_{(j)}.\nonumber\\
 \label{eq:C-alpha-Super}
\end{eqnarray}
Here, $ a_{0j}$ and  $b_{0j}$ stand for $a(u,z_j)$ and $b(u,z_j)$,
respectively.

\sect{$gl(2|2)$ Bethe vectors in the $F$-basis}

In this section, specializing to the superconductive $gl(2|2)$
electronic system, we will resolve its nested Bethe vectors in the
$F$-basis.

In the framework of the algebraic Bethe ansatz, the pseudo-vacuum
state is
\begin{eqnarray}
|0>=\otimes_{k=1}^N\left(\begin{array}{c}
  0\\0\\0\\1\end{array}\right)_{(k)}. \label{de:vacuum}
\end{eqnarray}
The Bethe vector of the model is then defined by
\begin{eqnarray}
\Phi_N(v_1,\ldots,v_{n_1})=\sum_{d_1\ldots d_{n_1}}
 (\phi^{(1)}_{n_1})^{d_1\ldots d_{n_1}}
 C_{d_1}(v_1)\ldots C_{d_{n_1}}(v_{n_1})|0>,
 \label{de:phi}
\end{eqnarray}
where $d_i=1,2,3$, $(\phi^{(1)}_{n_1})^{d_1\ldots d_{n_1}}$ is a
function of the spectral parameter $v_j$. In the algebraic Bethe
ansatz, $(\phi^{(1)}_{n_1})^{d_1\ldots d_{n_1}}$ is also
associated with the 3-dimensional nested Bethe vector
\begin{eqnarray}
\phi^{(1)}_{n_1}(v^{(1)}_{1},\ldots,v^{(1)}_{n_2})
 =\sum_{d_1\ldots d_{n_2}}
 (\phi^{(2)}_{n_2})^{d_1\ldots d_{n_2}}
 C^{(1)}_{d_1}(v^{(1)}_1)\ldots
 C^{(1)}_{d_{n_2}}(v^{(1)}_{{n_2}})|0>^{(1)}
 \label{de:phi-nested-1}
\end{eqnarray}
where $d_i=1,2$, $|0>^{(1)} $ is the nested pseudo-vacuum state
\begin{eqnarray}
|0>^{(1)}=\otimes_{k=1}^{n_1}\left(\begin{array}{c}
 0\\0\\1\end{array}\right), \label{de:vacuum-nested-1}
\end{eqnarray}
$(\phi^{(2)}_{n_2})^{d_1\ldots d_{n_2}}$ is a function of the
spectral parameter $v_j^{(1)}$, and $C^{(1)}$'s are the creation
operators of the nested $gl(2|1)$ system. Here $\phi^{(2)}_{n_2}$
is the second level nested Bethe vector associated with the
$gl(2)$ model.  As usual, the second level nested Bethe vector is
defined by
\begin{eqnarray}
\phi^{(2)}_{n_2}(v^{(2)}_{1},\ldots,v^{(2)}_{n_3})
 =
 C^{(2)}(v^{(2)}_1)\ldots
 C^{(2)}(v^{(2)}_{{n_3}})|0>^{(2)}, \label{de:phi-nested-2}
\end{eqnarray}
with
\begin{eqnarray} |0>^{(2)}=\otimes_{k=1}^{n_2}\left(\begin{array}{c}
  0\\1\end{array}\right)_{(k)}.  \label{de:vacuum-nested-2}
\end{eqnarray}

Applying the $gl(2|2)$ $F$-matrix, i.e. $m=2,n=2$ in (\ref{de:F}),
to the pseudo-vacuum state (\ref{de:vacuum}), we find it is
invariant. This is due to the fact that only terms whose roots all
equal to 4, in the definition expression of $F^{(1)}$, produce
non-zero results. Therefore the Bethe vector (\ref{de:phi}) in the
$F$-basis becomes
\begin{eqnarray}
\tilde\Phi_N(v_1,\ldots,v_{n_1})
 &\equiv& F_{1\ldots N}\Phi_N(v_1,\ldots,v_{n_1})\nonumber\\
 &=&\sum_{d_1\ldots d_{n_1}}(\phi^{(1)}_{n_1})^{d_1\ldots d_{n_1}}
 \tilde C_{d_1}(v_1)\ldots \tilde C_{d_{n_1}}(v_{{n_1}})|0>.
 \label{de:phi-F}
\end{eqnarray}

For the Bethe vector (\ref{de:phi-F}), one checks that the
spectral parameters of the system preserve the exchange symmetry
\begin{eqnarray}
\tilde\Phi_{N}(v_{\sigma(1)},\ldots,v_{\sigma(n_1)})
 = \tilde\Phi_{N}(v_1,\ldots,v_{n_1}),
  \label{eq:exchange}
\end{eqnarray}
while the creation operators satisfy the following commutation
relation \cite{Ess9211}
\begin{eqnarray}
\tilde C_{i}(u)\tilde C_{j}(v)
 =(-1)^{[i][j]}{1\over a(v,u)}
 \tilde C_{j}(v)\tilde C_{i}(u)
 -(-1)^{[j]}{b(v,u)\over a(v,u)}
 \tilde C_{j}(u)\tilde C_{i}(v).\label{eq:commu-cc}
\end{eqnarray}
These two properties mean that we may propose the following
special sequence of the creation operators with $p_1$ number of
$d_i=1$, $p_2-p_1$ number of $d_i=2$ and $n_1-p_2$ number of
$d_i=3$:
\begin{eqnarray}
 \tilde C_1(v_1)\ldots\tilde C_1(v_{p_1})
 \tilde C_2(v_{p_1+1})\ldots\tilde C_2(v_{p_2})
 \tilde C_3(v_{p_2+1})\ldots\tilde C_3(v_{n_1}).
 \label{eq:sequence-c123}
\end{eqnarray}
With the help of the commutation relation (\ref{eq:commu-cc}), we
may rewrite the sequence as
\begin{eqnarray}
 &&\tilde C_1(v_1)\ldots\tilde C_1(v_{p_1})
   \tilde C_2(v_{p_1+1})\ldots\tilde C_2(v_{p_2})
   \tilde C_3(v_{p_2+1})\ldots\tilde C_3(v_{n_1}) \nonumber\\
 &=&h(v_1,\ldots,v_{n_1})
 \tilde C_3(v_{p_2+1})\ldots\tilde C_3(v_{{n_1}})
 \tilde C_2(v_{p_1+1})\ldots\tilde C_2(v_{{p_2}})
 \nonumber \\ && %\quad\quad\quad\quad
 \times
 \tilde C_1(v_{1})\ldots
 \tilde C_1(v_{p_1})+\ldots\
 \label{eq:C3-C2-C1}
\end{eqnarray}
with
$$h(v_1,\ldots,v_{n_1})=\prod_{k=1}^{p_1}\prod_{l=p_1+1}^{{p_2}}
(-{1\over a(v_l,v_k)})\prod_{k=1}^{p_2} \prod_{l=p_2+1}^{{n_1}}
 ({1\over a(v_l,v_k)}).$$
Here the prefactor $h$ comes from the first term of
(\ref{eq:commu-cc}), and $``\ldots$" stands for the other terms
contributed by the second term. Considering the exchange symmetry
(\ref{eq:exchange}), one easily checks that the other terms
$``\ldots$" can be represented by
\begin{eqnarray}
C_3(v_{\sigma(p_2+1)})\ldots\tilde C_3(v_{\sigma(n_1)})
 \tilde C_2(v_{\sigma(p_1+1)})\ldots\tilde C_2(v_{\sigma(p_2)})
 %\nonumber \\ && %\quad\quad\quad\quad
 %\times
 \tilde C_1(v_{\sigma(1)})\ldots
 \tilde C_1(v_{\sigma(p_1)}),
\end{eqnarray}
where $\sigma\in {\cal S}_{n_1}$. Substituting
(\ref{eq:sequence-c123}) into the Bethe vector (\ref{de:phi-F}),we
then propose the following Bethe vector $\Phi_N^{(p_1,p_2)}$
corresponding to the quantum number $p_1$ and $p_2$:
\begin{eqnarray}
&&\tilde\Phi_N^{(p_1,p_2)}(v_1,\ldots,v_{n_1})
 \nonumber\\
 &=&(\phi^{(1)}_{n_1})^{1\ldots 12\cdots23\cdots 3}
 \tilde C_{1}(v_1)\ldots \tilde C_1(v_{p_1})
 \tilde C_{2}(v_{p_1+1})\ldots \tilde C_2(v_{p_2})
 %\nonumber\\ &&\times
 \tilde C_{3}(v_{p_2+1})\ldots \tilde C_3(v_{n_1})|0>
 \nonumber\\
 &=&{1\over p_1!(p_2-p_1)!(n_1-p_2)!}
 \sum_{\sigma\in{\cal S}_{n_1}}
 (\phi^{(1),\sigma}_{n_1})^{1\ldots 12\cdots23\cdots 3}
 \tilde C_{3}(v_{\sigma(p_2+1)})\ldots \tilde C_3(v_{\sigma(n_1)})
 \nonumber\\ &&\times
 \tilde C_{2}(v_{\sigma(p_1+1)})\ldots \tilde C_2(v_{\sigma(p_2)})
 \tilde C_{1}(v_{\sigma(1)})\ldots \tilde C_1(v_{\sigma(p_1)})|0>,
 \label{eq:phi-p1p2-C321}
\end{eqnarray}
where $\sigma$ in $\phi^{(1),\sigma}_{n_1}$ implies that the
permutation of the inhomogeneous parameters of the original nested
Bethe vector.

In view of (\ref{eq:C-alpha-Bose}) and (\ref{eq:C-alpha-Super}),
the $gl(2|2)$ creation operators $C_i (i=1,2,3)$ in the $F$-basis
take the form
\begin{eqnarray}
 \tilde C_3&=&\sum_{i=1}^{N}b_{0i}E^{34}_{(i)}\otimes_{j\ne i}
    \mbox{diag}\left(a_{0j},a_{0j},a_{0j}a_{ij}^{-1},1\right)_{(j)},
    \label{eq:C3-gl22}  \\
 \tilde C_2&=&\sum_{i=1}^{N}b_{0i}E^{24}_{(i)}\otimes_{j\ne i}
    \mbox{diag}\left(a_{0j},2a_{0j},a_{0j}a_{ij}^{-1},1\right)_{(j)}
    +\ldots\ ,   \label{eq:C2-gl22} \\
 \tilde C_1&=&\sum_{i=1}^{N}b_{0i}E^{14}_{(i)}\otimes_{j\ne i}
    \mbox{diag}\left(2a_{0j},a_{0j}a_{ij}^{-1},a_{0j}a_{ij}^{-1},1
    \right)_{(j)}+\ldots\ , \label{eq:C1-gl22}
\end{eqnarray}
where $``\ldots"$ stands for terms which contain more than one
generators, e.g. $E^{12}_{i}\otimes E^{24}_{j}$ etc.. Applying
$\tilde C_2$ and $\tilde C_1$ to the pseudo-vacuum state
(\ref{de:vacuum}), one finds that all other terms equal to zero.
Therefore, substituting (\ref{eq:C3-gl22})-(\ref{eq:C1-gl22}) into
(\ref{eq:phi-p1p2-C321}), we obtain
\begin{eqnarray}
&&\tilde\Phi_N^{(p_1,p_2)}(v_1,\ldots,v_{n_1})
 \nonumber\\
 &=&{1\over p_1!(p_2-p_1)!(n-p_2)!}
 \sum_{i_1<\ldots<i_{p_1}}\sum_{i_{p_1+1}<\ldots<i_{p_2}}
 \nonumber\\ && \times\sum_{i_{p_2+1}<\ldots<i_{n_1}}
 B^{(0)}_{n_1,(p_1,p_2)}(v_1,\ldots,v_{n_1};v^{(1)}_1,\ldots,
 v^{(1)}_{p_1},\ldots,
 v^{(1)}_{p_2}|z_{i_1},\ldots,z_{i_{n_1}})\nonumber\\ && \times
 \prod_{j={p_{2+1}}}^{{n_1}}E^{34}_{(i_j)}
 \prod_{j={p_{1+1}}}^{{p_2}}E^{24}_{(i_j)}
 \prod_{j=i_{1}}^{{p_1}}E^{14}_{(i_j)}|0>, \label{eq:phi-p1p2-phi1}
\end{eqnarray}
where $\{i_{k_a+1},i_{k_a+1},\ldots,i_{k_{a+1}}\}
\cap\{i_{k_b+1},i_{k_b+1},\ldots,i_{k_{b+1}}\}=\varnothing$
$(k_a\ne k_b \mbox{ and } k_a,k_b\in\{0 ,p_1,p_2\})$ and
\begin{eqnarray}
&&B^{(0)}_{n_1,(p_1,p_2)}(v_1,\ldots,v_{n_1};v^{(1)}_1,\ldots,
 v^{(1)}_{p_1},\ldots,
 v^{(1)}_{p_2}|z_{i_1},\ldots,z_{i_{n_1}})\nonumber\\
&=&\sum_{\sigma \in S_{n}}
  \prod_{k=1}^{p_1}\prod_{l=p_1+1}^{p_2}
    \left(-{a(v_{\sigma(l)},z_{i_k})
    \over a(v_{\sigma(l)},v_{\sigma(k)})}\right)
    \prod_{k=1}^{p_2} \prod_{l=p_2+1}^{{n_1}}
    \left({a(v_{\sigma(l)},z_{i_k})\over a(v_{\sigma(l)},v_{\sigma(k)})}\right)
      \nonumber\\ && \times
         (\phi^{(1),\sigma}_{n_1})^{1\ldots 12\cdots23\cdots 3}
     B_{n-p_2}(v_{\sigma(p_2+1)},\ldots,
                      v_{\sigma(n_1)}|z_{i_{p_1+1}},\ldots,z_{i_{n_1}})
                       \nonumber\\
&&\times B_{p_2-p_1}^*(v_{\sigma(p_1+1)},\ldots,
                      v_{\sigma(p_2)}|z_{i_{p_1+1}},\ldots,z_{i_{p_2}})
         B_{p_1}^*(v_{\sigma(1)},\ldots,
                       v_{\sigma(p_1)}|z_{i_{1}},\ldots,z_{i_{p_1}})\nonumber\\
\label{eq:B-gl22}
\end{eqnarray}
with
\begin{eqnarray}
 && B_p(v_1,\ldots,v_p|z_{1},\ldots,z_{p})=%\nonumber\\&=&
 \sum_{\sigma\in {\cal S}_p}
 \prod_{m=1}^p b(v_m,z_{\sigma(m)})
\prod_{l=m+1}^{p}{{a(v_{m},z_{\sigma(l)})}\over{a(z_{\sigma(m)},z_{\sigma(l)})}},
 \nonumber\\
 && B^*_p(v_1,\ldots,v_p|z_{1},\ldots,z_{p})=%\nonumber\\&=&
 \sum_{\sigma\in {\cal S}_p} \mbox{sign}(\sigma)
 \prod_{m=1}^p b(v_m,z_{\sigma(m)})
\prod_{l=m+1}^{p}2a(v_{m},z_{\sigma(l)}).\nonumber\\
\label{eq:B-B*}
\end{eqnarray}

In (\ref{eq:B-gl22}), we still need to determine the form of
$(\phi^{(1)}_{n_1})^{1\ldots 12\cdots23\cdots 3}$, which should be
evaluated in the original basis. Define
 $\tilde\phi^{(1)}_{n_1}\equiv F^{(1)}_{1\ldots
 n_1}\phi^{(1)}_{n_1}$. we now examine the relation between
 $\phi^{(1)}_{n_1}$ and $\tilde\phi^{(1)}_{n_1}$.

Write the nested pseudo-vacuum vector (\ref{de:vacuum-nested-1})
as
\begin{equation}
|0>^{(1)}\equiv |3\cdots 3>^{(1)},
\end{equation}
where the number of 3 is ${n_1}$. Then the nested Bethe vector
(\ref{de:phi-nested-1}) can be rewritten as
\begin{equation}
\phi^{(1)}_{n_1}(v_1^{(1)}\ldots v_{p_2}^{(1)})
 \equiv|\phi^{(1)}_{n_1}>
 =\sum_{d_1\ldots d_{n_1}}(\phi^{(1)}_{n_1})^{d_1\ldots d_{n_1}}|d_1\ldots d_{n_1}>^{(1)}.
 \label{eq:phi1-phi1}
\end{equation}
Acting the $gl(2|1)$ $F$-matrix $F^{(1)}$ from the left on the
above equation, we have
\begin{equation}
\tilde\phi^{(1)}_{n_1}(v_1^{(1)}\ldots v_{p_2}^{(1)})
 \equiv|\tilde\phi^{(1)}_{n_1}>=F^{(1)}|\phi^{(1)}_{n_1}>
 =\sum_{d_1\ldots d_{n_1}}(\tilde\phi^{(1)}_{n_1})^{d_1\ldots d_{n_1}}|d_1\ldots d_{n_1}>^{(1)}
 .
 \label{eq:phi1-phi1-F}
\end{equation}
It follows that
\begin{eqnarray}
  (\tilde\phi^{(1)}_{n_1})^{1\ldots 12\ldots23\ldots 3}
 &=&<1\ldots 12\ldots23\ldots 3|\tilde\phi^{(1)}_{n_1}>
  =<1\ldots 12\ldots23\ldots 3|F^{(1)}|\phi^{(1)}_{n_1}>\nonumber\\
 &=&<1\ldots 12\ldots23\ldots 3|\sum_{\sigma\in{\cal S}_{n_2}}
    \sum^{\quad\quad*}_{\alpha_{\sigma(1)}\ldots\alpha_{\sigma({n_1})}}
    \prod_{j=1}^{n_1} P_{\sigma(j)}^{\alpha_{\sigma(j)}}
    \nonumber\\ && \times
    S(c,\sigma,\alpha_\sigma)
    R^\sigma_{1\ldots {n_1}}
    |\phi^{(1)}_{n_1}> \label{eq:tildephi-phi-1}\nonumber\\
 &=&<1\ldots 12\ldots23\ldots 3|\left.\left\{
    \sum^{\quad\quad*}_{\alpha_{\sigma(1)}\ldots\alpha_{\sigma({n_1})}}\prod_{j=1}^{n_1}
    P_{\sigma(j)}^{\alpha_{\sigma(j)}}\right\}
    \right|_{\sigma=id} \nonumber\\ && \times
    S(c,\sigma,\alpha_\sigma)
    %R^{\sigma=id}_{1\ldots {n_2}}
    |\phi^{(1)}_{n_1}> \label{eq:tildephi-phi-2} \nonumber\\
 &=&t(c) <1\ldots 12\ldots23\ldots 3|\phi^{(1)}_{n_1}> %\nonumber\\
 =t(c)  (\phi^{(1)}_{n_1})^{1\ldots 12\ldots23\ldots 3}
 \label{eq:tildephi-phi-3}
\end{eqnarray}
with $t(c)=\prod_{j>i=1}^{p_1}(1+c_{ij})
    \prod_{j>i=p_1+1}^{p_2}(1+c_{ij})$.
It follows that $(\phi^{(1)}_{n_1})^{1\ldots 12\ldots23\ldots 3}$
may be computed by the $F$-transformed version
$(\tilde\phi^{(1)}_{n_1})^{1\ldots 12\ldots23\ldots 3}$.

In the following, we compute $(\phi^{(1)}_{n_1})^{1\ldots
12\cdots23\cdots 3}$ with the help of the nested $gl(2|1)$
$F$-basis. \\[5mm]
\underline {1. The first level nested $gl(2|1)$ Bethe vector}\\

Denoted by $F^{(1)}_{1\ldots n_1}$ the $n_1$-site $gl(2|1)$
$F$-matrix. Acting the $F$-matrix on (\ref{de:vacuum-nested-1}),
one finds that the nested $gl(2|1)$ pseudo-vacuum state is also
invariant.  Thus in the $F$-basis, the nested Bethe vector
(\ref{de:phi-nested-1}) becomes
\begin{eqnarray}
\tilde \phi^{(1)}_{n_1}(v^{(1)}_{1},\ldots,v^{(1)}_{n_2})
 =\sum_{d_1\ldots d_{n_2}}
 (\phi^{(2)}_{n_2})^{d_1\ldots d_{n_2}}
 \tilde C^{(1)}_{d_1}(v^{(1)}_1)\ldots
 \tilde C^{(1)}_{d_{n_2}}(v^{(1)}_{d_{n_2}})|0>^{(1)}.
 \label{de:phi-nested-1-F}
\end{eqnarray}

We may prove that the nested Bethe vector satisfies the following
exchange symmetry
\begin{eqnarray}
\tilde\phi^{(1)}_{n_1}(v^{(1)}_{\sigma(1)},\ldots,v^{(1)}_{\sigma(n_2)})
 ={1\over c^\sigma_{1\ldots {n_2}}}
  \tilde\phi^{(1)}_{n_1}(v^{(1)}_1,\ldots,v^{(1)}_{n_2}),
  \label{eq:exchange-nested}
\end{eqnarray}
where $c^\sigma_{1,\ldots,n}$ has the decomposition law
\begin{eqnarray}
 c^{\sigma'\sigma}_{1\ldots n}
 =c^{\sigma}_{\sigma'(1\ldots n)}
  c^{\sigma'}_{1\ldots n}\label{eq:c-cc}
\end{eqnarray}
with $c^{\sigma_i}_{1\ldots n}=c_{i\ i+1}\equiv c(v_i,v_{i+1})$
for an elementary permutation $\sigma_i$.

 This enable one to concentrate on a particularly simple term in the sum
(\ref{de:phi-nested-1-F}) of the form with $p_1$ number of $d_i=1$
and $n-p_1$ number of $d_j=2$:
\begin{eqnarray}
 \tilde C^{(1)}_1(v^{(1)}_1)\ldots\tilde C^{(1)}_1(v^{(1)}_{p_1})
 \tilde C^{(1)}_2(v^{(1)}_{p_1+1})\ldots\tilde C^{(1)}_2(v^{(1)}_{n_2}).
% \equiv g_{1\ldots 12\ldots 2}(v_1,\ldots,p_1,p_{1}+1,\ldots n_2).
     \label{eq:C1-C2}
\end{eqnarray}
The commutation relation between $C_i(v)$ and $C_j(u)$
\cite{Ess9211} in the $F$-basis becomes
\begin{eqnarray}
 \tilde C^{(1)}_i(v)\tilde C^{(1)}_j(u)
 &=&-{1\over a(u,v)}\tilde C^{(1)}_j(u)\tilde C^{(1)}_i(v)
    +{b(u,v)\over a(u,v)}\tilde C^{(1)}_j(v)\tilde C^{(1)}_i(u).
    \label{eq:commu-CC-F}
\end{eqnarray}
Then using (\ref{eq:commu-CC-F}), all $\tilde C_1^{(1)}$'s in
(\ref{eq:C1-C2}) can be moved to the right of all $\tilde
C_2^{(1)}$'s, yielding
\begin{eqnarray}
 &&\tilde C_1^{(1)}(v_1^{(1)})\ldots\tilde C_1^{(1)}(v_{p_1}^{(1)})
   \tilde C_2^{(1)}(v_{p_1+1}^{(1)})\ldots\tilde C_2^{(1)}(v_{n_2}^{(1)})
   = \nonumber\\
 &=&g(v_1^{(1)},\ldots,v_{n_2}^{(1)})
 \tilde C_2^{(1)}(v_{p_1+1}^{(1)})\ldots\tilde C_2^{(1)}(v_{{n_2}}^{(1)})
 \tilde C_1^{(1)}(v_{1}^{(1)})\ldots\tilde C_1^{(1)}(v_{p_1}^{(1)})+\ldots\ ,
 \label{eq:C2-C1}
\end{eqnarray}
where
$g(v_1^{(1)},\ldots,v_{n_2}^{(1)})=\prod_{k=1}^{p_1}\prod_{l=p_1+1}^{{n_2}}
(-{1/ a(v_l^{(1)},v_k^{(1)})})$ is the contribution from the first
term of (\ref{eq:commu-CC-F}). As before, the other terms
``$\ldots$" can be given with the help of the permutation operator
$\sigma\in {\cal S}_{n_2}$. Thus, we rewrite the nested Bethe
vector as
\begin{eqnarray}
 &&{\tilde\phi}_{n_1}^{{(1)},p_1}(v_1^{(1)},\ldots,v_{n_2}^{(1)})= \nonumber\\
 &=&{1\over p_1!({n_2}-p_1)!}\sum_{\sigma \in {\cal S}_{{n_2}}}
     c^{\sigma}_{1\ldots {n_2}}
    (\phi^{(2),\sigma}_{n_2})^{11\ldots 12\ldots 2}
    \prod_{k=1}^{p_1}\prod_{l=p_1+1}^{{n_2}}
    \left(-{1\over a(v^{(1)}_{\sigma(l)},v^{(1)}_{\sigma(k)})}\right)
    \nonumber\\
&&\times\;{\tilde C}_{2}^{(1)}(v^{(1)}_{\sigma(p_1 +1)})\ldots
{\tilde C}_{2}^{(1)}(v^{(1)}_{\sigma({n_2})}) {\tilde
C}_{1}^{(1)}(v^{(1)}_{\sigma(1)})\ldots {\tilde
C}_{1}^{(1)}(v^{(1)}_{\sigma(p_1)})\,|0>\, .\label{eq:Phi-C2C1}
\end{eqnarray}

From (\ref{eq:C-alpha-Bose}) and (\ref{eq:C-alpha-Super}), the
$gl(2|1)$ creation operators in the $F$-basis are given by
\begin{eqnarray}
 \tilde C_2&=&\sum_{i=1}^{N}b_{0i}E^{23}_{(i)}\otimes_{j\ne i}
    \mbox{diag}\left(a_{0j},2a_{0j},1\right)_{(j)} ,
    \label{eq:C2-gl21} \\
 \tilde C_1&=&\sum_{i=1}^{N}b_{0i}E^{13}_{(i)}\otimes_{j\ne i}
    \mbox{diag}\left(2a_{0j},a_{0j}a_{ij}^{-1},1
    \right)_{(j)}+\ldots\ , \label{eq:C1-gl21}
\end{eqnarray}
Substituting these expressions of $\tilde C_i$ into
(\ref{eq:Phi-C2C1}), we have
\begin{eqnarray}
&&{\tilde\phi}_{n_1}^{{(1)},p_1}(v_1^{(1)},\ldots,v_{n_2}^{(1)})\nonumber\\
 &=&{1\over p_1!({n_2}-p_1)!}
 \sum_{i_1<\ldots<i_{p_1}}\sum_{i_{p_1+1}<\ldots<i_{n_2}}
  B^{(1)}_{{n_2},p_1}(v_1^{(1)},\ldots,v_{{n_2}}^{(1)};v_{1}^{(2)},\ldots,
              v_{p_1}^{(2)}|v_{i_1},\ldots,v_{i_{{n_2}}})\nonumber\\
&&\times  \prod_{j={p_1+1}}^{{{n_2}}}
E^{23}_{(i_j)}\prod_{j={1}}^{{p_1}}E^{13}_{(i_j)}\,|0>,
\end{eqnarray}
where
$\{i_1,i_2,\ldots,i_{p_1}\}\cap\{i_{p_1+1},i_{p_1+2},\ldots,i_{n_2}\}=\varnothing$
and
\begin{eqnarray}
 &&B^{(1)}_{{n_2},p_1}(v_1^{(1)},\ldots,v_{{n_2}}^{(1)};v_{1}^{(2)},\ldots,
              v_{p_1}^{(2)}|v_{i_{1}},\ldots,v_{i_{{n_2}}})=\nonumber\\
 &=&\sum_{\sigma \in S_{{n_2}}}
    c^{\sigma}_{1\ldots {n_2}}
  \prod_{k=1}^{p_1}\prod_{l=p_1+1}^{{n_2}}
    \left(-{a(v^{(1)}_{\sigma(l)},v_{i_k})\over
           a(v_{\sigma(l)}^{(1)}-v^{(1)}_{\sigma(k)})}
           \right)
      \nonumber\\ && \times
         (\phi^{(2),\sigma}_{n_2})^{11\ldots 12\ldots 2}
         B_{{n_2}-p_1}^*(v^{(1)}_{\sigma(p_1+1)},\ldots,
                      v^{(1)}_{\sigma({n_2})}|v_{i_{p_1+1}},\ldots,v_{i_{{n_2}}})
                       \nonumber\\
&&\times
         B_{p_1}^*(v_{\sigma(1)}^{(1)},\ldots,
                       v^{(1)}_{\sigma(p_1)}|v_{i_{1}},\ldots,v_{i_{p_1}}).\nonumber\\
\label{eq:B1-gl21}
\end{eqnarray}

Denote by $F^{(2)}$ the second level nested $gl(2)$ $F$-matrix.
Applying the $F$-matrix to $\phi^{(2)}_{n_2}$, one obtains
\begin{equation}
 (\phi^{(2)}_{n_2})^{11\ldots 12\ldots 2}
 =\left(t'(c)\right)^{-1}(\tilde\phi^{(2)}_{n_2})^{11\ldots 12\ldots 2}
\end{equation}
with  $t'(c)=\prod_{j>i=1}^{p_1}(1+c_{ij})
    \prod_{j>i=p_1+1}^{{n_2}}(1+c_{ij})$. Therefore as before, the
$(\phi^{(2),\sigma}_{n_2})^{11\ldots 12\ldots 2}$ in
(\ref{eq:B1-gl21}) can be determined in the $gl(2)$
$F$-basis.\\[5mm]
\underline {2. The second level nested $gl(2)$ Bethe vector}\\

The $m=2,$ and $n=0$ limit of (\ref{de:F}) gives the $gl(2)$
$F$-matrix $F^{(2)}$. In this $F$-basis, the $n_2$-site simple
generator $E^{12}$ is given by
\begin{eqnarray}
\tilde E^{(12)}=\sum_{i=1}^{n_2}E^{12}_{(i)}\otimes_{j\ne i}
 \mbox{diag}\left(2,(2a_{ij})^{-1}\right)_{(j)},
\end{eqnarray}
the diagonal element $D^{(2)}(u)$ of the $gl(2)$ monodromy matrix
is
\begin{eqnarray}
\tilde
D^{(2)}(u)=\otimes_{i=1}^{n_2}\mbox{diag}(a_{0i},c_{0i})_{(0)},
\end{eqnarray}
and the creation operator $C^{(2)}(u)$ becomes
\begin{eqnarray}
\tilde C^{(2)}(u)=-\sum_{i=1}^{n_2}b_{0i}E^{12}_{(i)}
 \otimes_{j\ne i}
 \mbox{diag}\left(2a_{0j},(2a_{ij})^{-1}c_{0j}\right)_{(j)}.
 \label{eq:C-nested-2}
\end{eqnarray}

Applying $F^{(2)}$ to the nested Bethe vector $\phi^{(2)}_{n_2}$,
we obtain
\begin{eqnarray}
  \tilde\phi^{(2)}_{n_2}(v_1^{(2)},\ldots, v_{n_3}^{(2)})
 &\equiv& F^{(1)}_{1\ldots {n_2}}
          \phi^{(2)}(v_1^{(2)},\ldots, v_{n_3}^{(2)}) \nonumber\\
 &=&s_{n_3}(c)\tilde C^{(2)}(v^{(2)}_1)
\tilde C^{(2)}(v^{(2)}_2)\ldots
           \tilde C^{(2)}(v^{(2)}_{n_3})|0>^{(2)},
\label{de:phi-nested-2-F}
\end{eqnarray}
where $s_{n}(c)=\prod_{j>i=1}^{n}(1+c_{ij})$ is from the action of
$F^{(2)}$ on the nested pseudo-vacuum state $|0>^{(1)}$. With the
help of (\ref{eq:C-nested-2}), the $F$-transformed nested Bethe
vector is given by
\begin{eqnarray}
 &&\tilde \phi^{(2)}_{n_2}(v_1^{(2)},\ldots,v_{n_3}^{(2)})
   =s_{n_3}(c)\tilde{C}^{(2)}(v_1^{(2)})\ldots
    \tilde{C}^{(2)}(v_{n_3}^{(2)})
   \,|0>^{(2)}\nonumber\\
&=&s_{n_3}(c)\sum_{i_1<\ldots< i_{{n_3}}}
B_{{n_3}}^{(2)}(v_1^{(2)},\ldots,v_{n_3}^{(2)}|v^{(1)}_{i_1},
  \ldots,v^{(1)}_{i_{{n_3}}})
E_{(i_1)}^{12}\ldots E_{(i_{n_3})}^{12}\,|0>^{(2)}\;,
\label{Psi_2}
\end{eqnarray}
where
\begin{eqnarray}
&& B^{(2)}_{n_3}(v_1^{(2)},\ldots,
     v_{n_3}^{(2)}|v^{(1)}_{2},\ldots,v^{(1)}_{{n_3}})
\nonumber\\
 &=&\sum_{\sigma\in {\cal S}_n}\prod_{k=1}^{n_3}
 \left(-b(v^{(2)}_k,v^{(1)}_{\sigma(k)}\right)
%\nonumber\\ && \times
  \prod^{n_2}_{j\ne \sigma(1),\ldots,\sigma(k)}
  {c(v^{(2)}_k,v^{(1)}_{j})\over
   2a(v^{(1)}_{\sigma(k)},v_{j}^{(1)})}
 %  \nonumber\\ &&\times
 \prod_{l=k+1}^m
 2a(v^{(2)}_k,v_{\sigma(l)}^{(1)}). \label{eq:B2-gl2}
\end{eqnarray}

Here we note that if we exchange the spectral parameter in
(\ref{de:phi-nested-2-F}), the $gl(2)$ Bethe vector is invariant.
Therefore substituting the representation of $\tilde \phi^{(2)}$,
i.e.  $B^{(2)}$,  into (\ref{eq:B1-gl21}), we may rewrite the
nested $gl(2|1)$ Bethe vector as
\begin{eqnarray}
&&{\tilde\phi}_{n_1}^{{(1)},p_1}(v_1^{(1)},\ldots,v_{n_2}^{(1)})\nonumber\\
 &=&{s_{p_1}(c)\over p_1!({n_2}-p_1)!}
 \sum_{i_1<\ldots<i_{p_1}}\sum_{i_{p_1+1}<\ldots<i_{n_2}}
  B^{(1)}_{{n_2},p_1}(v_1^{(1)},\ldots,v_{{n_2}}^{(1)};v_{1}^{(2)},\ldots,
              v_{p_1}^{(2)}|v_{i_1},\ldots,v_{i_{{n_2}}})\nonumber\\
&&\times  \prod_{j={p_1+1}}^{{{n_2}}}
E^{23}_{(i_j)}\prod_{j={1}}^{{p_1}}E^{13}_{(i_j)}\,|0>
\end{eqnarray}
with
\begin{eqnarray}
 &&B^{(1)}_{{n_2},p_1}(v_1^{(1)},\ldots,v_{{n_2}}^{(1)};v_{1}^{(2)},\ldots,
              v_{p_1}^{(2)}|v_{i_{1}},\ldots,v_{i_{{n_2}}})\nonumber\\
 &=&\sum_{\sigma \in S_{{n_2}}}
    c^{\sigma}_{1\ldots {n_2}}
\left(\prod_{\epsilon=0}^{1}
    \prod_{\sigma(j)>\sigma(i)=p_{\epsilon}+1}^{p_{\epsilon+1}}
    (1+c_{\sigma(i)\sigma(j)})\right)^{-1}
    \nonumber\\&&\times
  \prod_{k=1}^{p_1}\prod_{l=p_1+1}^{{n_2}}
    \left(-{a(v^{(1)}_{\sigma(l)},v_{i_k})\over
           a(v_{\sigma(l)}^{(1)},v^{(1)}_{\sigma(k)})}
           \right)
      \nonumber\\ && \times
       B^{(2)}_{p_1}(v_{1}^{(2)},\ldots,v_{p_1}^{(2)}|v^{(1)}_{\sigma(1)},\ldots,
         v^{(1)}_{\sigma(p_1)}) \nonumber\\ &&\times
       B_{{n_2}-p_1}^*(v^{(1)}_{\sigma(p_1+1)},\ldots,
         v^{(1)}_{\sigma({n_2})}|v_{i_{p_1+1}},\ldots,v_{i_{{n_2}}})
         \nonumber\\ &&\times
       B_{p_1}^*(v_{\sigma(1)}^{(1)},\ldots,
         v^{(1)}_{\sigma(p_1)}|v_{i_{1}},\ldots,v_{i_{p_1}}).\nonumber\\
\label{B1}
\end{eqnarray}
\\[4mm]

Having resolved the nested $gl(2|1)$ Bethe vector, we now go back
the Bethe vectors of $gl(2|2)$ electronic model. By the exchange
symmetry of the $gl(2|1)$ Bethe vector (\ref{eq:exchange-nested}),
we represent the Bethe vector (\ref{eq:phi-p1p2-phi1}) in the
$F$-basis by
\begin{eqnarray}
&&\tilde\Phi_N^{(p_1,p_2)}(v_1,\ldots,v_{n_1})
 \nonumber\\
 &=&{s_{p_1}(c)\over (p_1!)^2((p_2-p_1)!)^2(n_1-p_2)!}%
 \sum_{i_1<\ldots<i_{p_1}}\sum_{i_{p_1+1}<\ldots<i_{p_2}}
 \nonumber\\ && \times\sum_{i_{p_2+1}<\ldots<i_{n_1}}
 B^{(0)}_{n_1,(p_1,p_2)}(v_1,\ldots,v_{n_1};v^{(1)}_1,\ldots,
 v^{(1)}_{p_1},\ldots,
 v^{(1)}_{p_2}|z_{i_1},\ldots,z_{i_{n_1}})\nonumber\\ && \times
 \prod_{j={p_{2}+1}}^{{n_1}}E^{34}_{(i_j)}
 \prod_{j={p_{1}+1}}^{{p_2}}E^{24}_{(i_j)}
 \prod_{j={1}}^{{p_1}}E^{14}_{(i_j)}|0> \label{eq:phi-p1p2-phi1}
\end{eqnarray}
with
\begin{eqnarray}
&&B^{(0)}_{n_1,(p_1,p_2)}(v_1,\ldots,v_{n_1};v^{(1)}_1,\ldots,
 v^{(1)}_{p_2}|z_{i_1},\ldots,z_{i_{n_1}})\nonumber\\
&=&\sum_{\sigma \in S_{n_1}} \left(\prod_{\epsilon=0}^{1}
    \prod_{\sigma(j)>\sigma(i)=p_{\epsilon}+1}^{p_{\epsilon+1}}
    (1+c_{\sigma(i)\sigma(j)})\right)^{-1}
      \nonumber\\ && \times
 %  \left(c^\sigma_{1\ldots n_1}\right)^{-1}
  \prod_{k=1}^{p_1}\prod_{l=p_1+1}^{p_2}
    \left(-{a(v_{\sigma(l)},z_{i_k})
    \over a(v_{\sigma(l)},v_{\sigma(k)})}\right)
  \prod_{k=1}^{p_2} \prod_{l=p_2+1}^{{n_1}}
    {a(v_{\sigma(l)},z_{i_k})\over a(v_{\sigma(l)},v_{\sigma(k)})}
      \nonumber\\ && \times
    B^{(1)}_{{p_2},p_1}(v_1^{(1)},\ldots,v_{{p_2}}^{(1)};v_{1}^{(2)},
    \ldots,v_{p_1}^{(2)}|v_{\sigma(i_{1})},\ldots,v_{\sigma(i_{p_2})})
    \nonumber\\ && \times
    B_{n-p_2}(v_{\sigma(p_2+1)},\ldots,
                      v_{\sigma(n_1)}|z_{i_{p_1+1}},\ldots,z_{i_{n_1}})
                       \nonumber\\
&&\times
    B_{p_2-p_1}^*(v_{\sigma(p_1+1)},\ldots,
                      v_{\sigma(p_2)}|z_{i_{p_1+1}},\ldots,z_{i_{p_2}})
   \nonumber\\ && \times
    B_{p_1}^*(v_{\sigma(1)},\ldots,
                       v_{\sigma(p_1)}|z_{i_{1}},\ldots,z_{i_{p_1}}).\nonumber\\
\end{eqnarray}

\sect{The resolution of the $gl(m|n)$ nested Bethe vectors in the
$F$-basis}

In this section, we generalize the results in the previous section
to the $gl(m|n)$ supersymmetric model. The procedure is similar to
that of the $gl(2|2)$ case. Here we only give the final results.

Associated with the $(m+n)$-dimensional $gl(m|n)$ representation
space, we have the orthogonal states $|j>\ (j=1,2,\ldots, m+n)$
defined by
\begin{eqnarray}
|1>=\left(\begin{array}{c} 1\\0\\ \vdots \\ 0\end{array}\right)
\quad,\ldots, \quad
 |m+n>=\left(\begin{array}{c} 0\\0\\ \vdots \\1
 \end{array}\right).
\end{eqnarray}

Consider the  $gl(m|n)$ Bethe state vectors with quantum numbers
$p_1,\ p_2-p_1,\ldots,\\ p_{m+n-2}-p_{m+n-3},\ n_1-p_{m+n-2}$, and
$N-n_1$, which label the numbers of state $|1>$, $|2>$, $\ldots,$
$|m+n>$ in the Bethe state, respectively. Define the pseudo-vacuum
state of the $N$-site system
$$|0>=\otimes_{k=1}^{N}|m+n>_{(k)}.$$ Then the $gl(m|n)$ Bethe
vector in the $F$-basis corresponding to the special quantum
numbers, $\Phi_N^{(p_1,\ldots,p_{m+n-2})}(v_1,\ldots,v_{n_1})$, is
given by
%i) for $n>2$
\begin{eqnarray}
&&\tilde\Phi_N^{(p_1,\ldots,p_{m+n-2})}(v_1,\ldots,v_{n_1})
 \nonumber\\
 &=&{\prod_{\alpha=1}^{m-1}s_{p_\alpha}(c)p_1!\over
     \prod_{\alpha=1}^{m+n-1}
     ((p_{\alpha}-p_{\alpha-1})!)^{m+n-\alpha}}
%  \nonumber\\ && \times
\sum_{i_1<\ldots<i_{p_1}}\sum_{i_{p_1+1}<\ldots<i_{p_2}}\ldots
\sum_{i_{p_{m+2-2}+1}<\ldots<i_{n_1}}
 \nonumber\\ && \times
 B^{(0)}_{n_1,(p_1,\ldots,p_{m+n-2})}
 (v_1,\ldots,v_{n_1};v^{(1)}_1,\ldots,
 v^{(1)}_{p_1},\ldots,
 v^{(1)}_{p_{m+n-2}}|z_{i_1},\ldots,z_{i_{n_1}})\nonumber\\ && \times
 \prod_{\alpha=1}^{m+n-1}
 \prod_{j={p_{m+n-\alpha-1}+1}}^{p_{m+n-\alpha}}
 E^{m+n-\alpha\ m+n}_{(i_j)}|0>,
 \nonumber\\
\end{eqnarray}
where $\{i_{k_a+1},i_{k_a+2},\ldots,i_{k_{a+1}}\}\cap
\{i_{k_b+1},i_{k_b+2},\ldots,i_{k_{b+1}}\}=\varnothing\,$ $(k_a\ne
k_b \mbox{ and } k_a,k_b\in\{p_0,p_1\ldots,p_{m+n-2}\})$, for
$m=1$,
\begin{eqnarray}
&&B^{(0)}_{n_1,(p_1,\ldots,p_{m+n-2})}
 (v_1,\ldots,v_{n_1};v^{(1)}_1,\ldots,
 v^{(1)}_{p_1},\ldots,
 v^{(1)}_{p_{m+n-2}}|z_{i_1},\ldots,z_{i_{n_1}}) \nonumber\\
&=&B^{*}_{p_1}
     \left(v^{(n-1)}_{1},\ldots,
      v^{(n-1)}_{p_1}|
      v^{(n-2)}_{\sigma^{2}(1)},\ldots,
      v^{(n-2)}_{\sigma^{2}(p_1)}\right)
      \nonumber\\ &&\times
  \prod_{\beta=2}^{n}\left[
   \sum_{\sigma^{\beta} \in {\cal S}_{p_\beta}}
   \left(
    \prod_{\sigma^\beta(j)>\sigma^\beta(i)=1}^{p_{1}}
    (1+c_{\sigma^\beta(i)\sigma^\beta(j)})\right)^{-1}\right.
%  \nonumber\\ && \times
  \prod_{\alpha=1}^{n-1}
  \prod_{k=1}^{p_\alpha} \prod_{l=p_\alpha+1}^{p_{\alpha+1}}
    {a\left(v^{(n-\beta)}_{\sigma^{\beta}(l)},
     v^{(n-1-\beta)}_{\sigma^{\beta+1}(k_{\alpha})}\right)
    \over a\left(v^{(n-\beta)}_{\sigma^{\beta}(l)},
    v^{(n-\beta)}_{\sigma^{\beta}(k)}\right)}
     \nonumber\\ &&\quad \times
  \prod_{\theta=1}^{\beta} \left(
    B^*_{p_1}
     \left(v^{(n-\theta)}_{\sigma^{\theta}(1)},\ldots,
      v^{(n-\theta)}_{\sigma^{\theta}(p_1)}|
      v^{(n-1-\theta)}_{\sigma^{\theta+1}(1)},\ldots,
      v^{(n-1-\theta)}_{\sigma^{\theta+1}(p_{1})}\right)\right.
      \nonumber\\ && \quad\times \left.\left.
      \prod_{\gamma=2}^{\theta}
    B_{p_\gamma-p_{\gamma-1}}
     \left(v^{(n-\theta)}_{\sigma^{\theta}(p_{\gamma-1}+1}),\ldots,
      v^{(n-\theta)}_{\sigma^{\theta}(p_{\gamma}+1)}|
      v^{(n-1-\theta)}_{\sigma^{\theta+1}(p_{\gamma-1}+1)},\ldots,
      v^{(n-1-\theta)}_{\sigma^{\theta+1}(p_{\gamma}+1)}\right)
      \right)\right]
      \nonumber\\
\end{eqnarray}
and for $m>1$,
\begin{eqnarray}
&&B^{(0)}_{n_1,(p_1,\ldots,p_{m+n-2})}
 (v_1,\ldots,v_{n_1};v^{(1)}_1,\ldots,
 v^{(1)}_{p_1},\ldots,
 v^{(1)}_{p_{m+n-2}}|z_{i_1},\ldots,z_{i_{n_1}}) \nonumber\\
&=&B^{**}_{p_1}
     \left(v^{(m+n-2)}_{1},\ldots,
      v^{(m+n-2)}_{p_1}|
      v^{(m+n-3)}_{\sigma^{2}(1)},\ldots,
      v^{(m+n-3)}_{\sigma^{2}(p_1)}\right)
      \nonumber\\ &&\times
        \prod_{\beta=2}^{m-1} \left[
  \sum_{\sigma^{\beta} \in {\cal S}_{p_\beta}}\left(
    \prod_{\epsilon=0}^{\beta-1}
    \prod_{\sigma^\beta(j)>\sigma^\beta(i)=p_{\epsilon}+1}^{p_{\epsilon+1}}
    (1+c_{\sigma^\beta(i)\sigma^\beta(j)})\right)^{-1}
  \right.\nonumber\\ &&\quad \times
  \prod_{\alpha=1}^{\beta-1}
  \prod_{k=1}^{p_\alpha}\prod_{l=p_\alpha+1}^{p_{\alpha}}
    {c\left(v^{(m+n-1-\beta)}_{\sigma^{\beta}(l)},
     v^{(m+n-1-\beta)}_{\sigma^{\beta}(k)}\right)
    a\left(v^{(m+n-1-\beta)}_{\sigma^{\beta}(l)},
     v^{(m+n-2-\beta)}_{\sigma^{\beta+1}(k_\alpha)}\right)
    \over a\left(v^{(m+n-1-\beta)}_{\sigma^{\beta}(l)},
     v^{(m+n-1-\beta)}_{\sigma^{\beta}(k)}\right)}
          \nonumber\\ &&\quad \times \left.
 \prod_{\theta=2}^{\beta}\prod_{\alpha=1}^{\theta}
    B^{**}_{p_\alpha-p_{\alpha-1}}
     \left(v^{(m+n-1-\theta)}_{\sigma^{\theta}(p_{\alpha-1}+1)},\ldots,
      v^{(m+n-1-\theta)}_{\sigma^{\theta}(p_{\alpha})}|
      v^{(m+n-2-\theta)}_{\sigma^{\theta+1}(p_{\alpha-1}+1)},\ldots,
      v^{(m+n-2-\theta)}_{\sigma^{\theta+1}(p_{\alpha})}\right)\right]
      \nonumber\\ && \times
\prod_{\beta=m}^{m+n-1}\left[
   \sum_{\sigma^{\beta} \in {\cal S}_{p_\beta}}
   \mbox{Exp}\left\{\delta_{\beta,m}\ln
    c_{1,\ldots p_{\beta}}^{\sigma^\beta}\right\}\left(
    \prod_{\epsilon=0}^{m-1}
    \prod_{\sigma^\beta(j)>\sigma^\beta(i)=p_{\epsilon}+1}^{p_{\epsilon+1}}
    (1+c_{\sigma^\beta(i)\sigma^\beta(j)})\right)^{-1}\right.
  \nonumber\\ && \quad\times
  \prod_{\alpha=1}^{m-1}
  \prod_{k=1}^{p_\alpha}\prod_{l=p_\alpha+1}^{p_{\alpha+1}}
    \left(-{a\left(v^{(m+n-1-\beta)}_{\sigma^{\beta}(l)},
     v^{(m+n-2-\beta)}_{\sigma^{\beta+1}(k_\alpha)}\right)
    \over a\left(v^{(m+n-1-\beta)}_{\sigma^{\beta}(l)},
    v^{(m+n-1-\beta)}_{\sigma^{\beta}(k)}\right)}\right)
     \nonumber\\ &&\quad  \times
  \prod_{\alpha=m}^{m+n-2}
  \prod_{k=1}^{p_\alpha} \prod_{l=p_\alpha+1}^{p_{\alpha+1}}
    {a\left(v^{(m+n-1-\beta)}_{\sigma^{\beta}(l)},
     v^{(m+n-2-\beta)}_{\sigma^{\beta+1}(k_{\alpha})}\right)
    \over a\left(v^{(m+n-1-\beta)}_{\sigma^{\beta}(l)},
    v^{(m+n-1-\beta)}_{\sigma^{\beta}(k)}\right)}
     \nonumber\\ &&\quad \times
  \prod_{\theta=m}^{\beta} \left(\prod_{\alpha=1}^{m}
    B^*_{p_\alpha-p_{\alpha-1}}
     \left(v^{(m+n-1-\theta)}_{\sigma^{\theta}(p_{\alpha-1}+1)},\ldots,
      v^{(m+n-1-\theta)}_{\sigma^{\theta}(p_{\alpha})}|
      v^{(m+n-2-\theta)}_{\sigma^{\theta+1}(p_{\alpha-1}+1)},\ldots,
      v^{(m+n-2-\theta)}_{\sigma^{\theta+1}(p_{\alpha})}\right)\right.
      \nonumber\\ && \quad\times \left.\left.
      \prod_{\gamma=m+1}^{\theta}
    B_{p_\gamma-p_{\gamma-1}}
     \left(v^{(m+n-1-\theta)}_{\sigma^{\theta}(p_{\gamma-1}+1)},\ldots,
      v^{(m+n-1-\theta)}_{\sigma^{\theta}(p_{\gamma})}|
      v^{(m+n-2-\theta)}_{\sigma^{\theta+1}(p_{\gamma-1}+1)},\ldots,
      v^{(m+n-2-\theta)}_{\sigma^{\theta+1}(p_{\gamma})}\right)
      \right)\right]
      \nonumber\\
\end{eqnarray}
with the conventions $p_0=0,\ p_{m+n-1}= n_1,\ v^{(0)}=v,\
v^{(-1)}= u,$ $\sigma^{m+n}=1,$ and $B^{**}_{n}(v_1^{(l)},\ldots,
     v_{n}^{(l)}|v^{(l-1)}_{1},\ldots,v^{(l-1)}_{{n}})=
 B^{(2)}_{n}(v_1^{(l)},\ldots,
     v_{n}^{(l)}|v^{(1)}_{l-1},\ldots,v^{(l-1)}_{{n}})$.
\\[6mm]

\section{Discussions}
In this paper, we have constructed the factorizing $F$-matrices
for the $gl(m|n)$-invariant fermion model. In the basis provided
by the $F$-matrix (the $F$-basis), the monodromy matrix and the
creation operators take completely symmetric forms. We moreover
have obtained a symmetric representation of the Bethe vector of
the system.

Authors in \cite{Korepin99} derived a formula that expresses the
local spin and field operators of fundamental graded models in
terms of the elements of the monodromy matrix. In particular they
reconstructed the local operators ($E^{ij}$) in terms of operators
figuring in the $gl(m|n)$ monodromy matrix. This together with the
results of the present paper in the $F$-basis should enable one to
get the exact representations of form factors and correlation
functions of the supersymmetric fermion models. These are under
investigation and results will be reported elsewhere.\\[5mm]
{\bf Acknowledgements:} This work was financially supported by the
Australia Research Council. S.Y. Zhao has also been supported by
the UQ Postdoctoral Research Fellowship.

 \end{document}